\documentclass[journal=jpcafh,manuscript=article]{achemso}

\usepackage{amsmath}
\usepackage{hyperref}

\setcitestyle{numbers,square}

\SectionNumbersOn

\title{Polyradical character and spin frustration in fullerene
  molecules:\\An {\em ab initio} non-collinear Hartree--Fock study}

\author{Carlos A. Jim\'enez-Hoyos}
\email{jimenez.hoyos@gmail.com}
\affiliation{Department of Chemistry, Rice University, Houston, TX
  77005}

\author{R. Rodr\'iguez-Guzm\'an}
\affiliation{Department of Chemistry, Rice University, Houston, TX
  77005}
\alsoaffiliation{Department of Physics and Astronomy, Rice University,
  Houston, TX 77005}

\author{Gustavo E. Scuseria}
\affiliation{Department of Chemistry, Rice University, Houston, TX
  77005}
\alsoaffiliation{Department of Physics and Astronomy, Rice University,
  Houston, TX 77005}

\date{\today}

\begin{document}

\begin{abstract}
Most {\em ab initio} calculations on fullerene molecules have been
carried out based on the paradigm of the H\"uckel model. This is
consistent with the restricted nature of the independent-particle
model underlying such calculations, even in single-reference-based
correlated approaches. On the other hand, previous works on some of
these molecules using model Hamiltonians have clearly indicated the
importance of short-range inter-atomic spin-spin correlations. In this
work, we consider {\em ab initio} non-collinear Hartree--Fock (HF)
solutions for representative fullerene systems: the bowl, cage, ring,
and pentagon isomers of C$_{20}$, and the larger C$_{30}$, C$_{36}$,
C$_{60}$, C$_{70}$, and C$_{84}$ fullerene cages. In all cases but the
ring we find that the HF minimum corresponds to a truly non-collinear
solution with a torsional spin density wave. Optimized geometries at
the generalized HF (GHF) level lead to fully symmetric structures,
even in those cases where Jahn-Teller distortions have been previously
considered. The nature of the GHF solutions is consistent with the
$\pi$-electron space becoming polyradical in nature: each $p$-orbital
remains effectively singly occupied. The spin frustration, induced by
the pentagon rings in an otherwise anti-ferromagnetic background, is
minimized at the HF level by aligning the spins in non-collinear
arrangements. The long-range magnetic ordering observed is reminiscent
of the character of broken symmetry HF solutions in polyacene systems.
\end{abstract}

\section{Introduction}

The discovery of fullerenes \cite{Kroto-1985,Heath-1985} and their
isolation in macroscopic quantities \cite{Kratschmer-1990} has
attracted interest in their electronic structure due to their unusual
topology and curvature. C$_{60}$ was expected to be the first example
of a spherical aromatic molecule, but its properties turned out not to
be consistent with this perspective. In particular, C$_{60}$ and other
fullerene compounds have chemical properties more similar to reactive
alkene molecules than to aromatic systems \cite{Taylor}.

Based on the inherent limitations conferred by the typical sizes of
fullerene molecules, {\em ab initio} quantum chemical calculations on
them are typically performed using a restricted formalism (see, {\em
  e.g.}, Refs. \cite{Scuseria-1996,Paulus-2004}) either based on
Hartree--Fock (HF) or density functional approximations. There has
been little reason to question this strategy since some of the most
common, stable fullerene molecules are non-magnetic and display a
large HOMO-LUMO gap. On the other hand, it has sporadically been noted
that some particular fullerene isomers (those with degeneracies in a
H\"uckel-type approach) possess large multi-reference character. It
was only recently reported by St\"uck {\em et al.} \cite{Stuck-2011}
that the restricted HF (RHF) solution for C$_{60}$ is unstable, {\em
  i.e.}, there exists a broken-symmetry solution which is lower in
energy. For the $D_{6h}$ configuration of C$_{36}$, Varganov and
co-workers \cite{Varganov-2002} had similarly reported a triplet
instability.

St\"uck {\em et al.} were seemingly unaware that the triplet
instability of C$_{60}$ was originally reported by Sheka
\cite{Sheka-2004} in semi-empirical calculations. In a series of works
\cite{Sheka-2004,Sheka-2006,Sheka-2007,Sheka-2011,Sheka}, Sheka and
collaborators have interpreted the physical and chemical properties of
fullerenes in terms of the unrestricted HF (UHF) solutions. She has
advocated an {\em odd electron} (or polyradical) interpretation of the
$\pi$-electron bonding in fullerenes, as opposed to the electron-gas
picture obtained within a H\"uckel-type restricted formalism. She has
also described C$_{60}$-based binary systems in terms of the
donor--acceptor interactions between the corresponding species
\cite{Sheka-2004b,Sheka-2007c}. Indeed, our interpretation of the
bonding in fullerenes in this contribution follows the lines already
elaborated by Sheka since 2004. The main difference lies in the use of
non-collinear spins, which leads to a fully symmetric $I_h$
configuration for C$_{60}$, consistent with a single $^{13}$C-NMR peak
\cite{Taylor-1990}, as opposed to the lesser symmetry present in UHF.

At this point we want to mention that the non-collinear arrangement of
spins in C$_{60}$ was originally described by Coffey and Trugman
\cite{Coffey-1992} in the framework of the classical Heisenberg
Hamiltonian solution. Konstantinidis
\cite{Konstantinidis-2005,Konstantinidis-2007} has more recently
discussed the unconventional magnetic properties developed in spin
lattices with icosahedral symmetry, such as that of C$_{60}$. Several
papers
\cite{Bergomi-1993,Joyes-1993,Scalettar-1993,Willaime-1993,Krivnov-1994,Sheng-1994,Srinivasan-1996,Srinivasan-1996b,Ojeda-1999,Flocke-2000,Lin-2007b}
studied C$_{60}$ using the Hubbard \cite{Hubbard-1963} and
Pariser-Parr-Pople (PPP) \cite{Pariser-1953,Pople-1953} model
Hamiltonians following Coffey and Trugman's original work. These
emphasized the intermediate-to-strongly correlated nature of the
physical system. We note that fully non-collinear HF solutions
\cite{Bergomi-1993,Ojeda-1999} to the Hubbard Hamiltonian displayed,
for sufficiently large $U/t$, the same type of spin arrangements
observed in the classical Heisenberg description. It remained unclear,
however, whether a full-electron calculation would lead to the same
type of mean-field solutions. In particular, Willaime and Falicov
\cite{Willaime-1993} observed that inclusion of Ohno-type screening
lead to the normal paramagnetic state in PPP calculations with
physical parameters.

The model Hamiltonian results stressed the role of anti-ferromagnetic
spin-spin correlations in fullerene lattices. It is crucial to stress
that a restricted-type solution is incompatible with the importance of
spin-spin correlations. In particular, RHF overemphasizes bond
alternation (or bond localization) and no spin ordering beyond that in
the (localized) $\pi$-bond. Quantum Monte Carlo
\cite{Scalettar-1993,Krivnov-1994,Srinivasan-1996} results in a
Hubbard framework have determined that spin-spin correlations in
C$_{60}$ are mid-ranged in nature (3-4 bonds), as opposed to the
longer correlation length predicted by a mean-field broken symmetry
treatment \cite{Bergomi-1993}.

We shall also mention the work of a few quantum chemists in trying to
understand the nature of the chemical bonding in fullerene molecules
from a valence bond treatment of the Heisenberg Hamiltonian (see
Ref. \cite{Schmalz-2002} and references therein). These studies have
helped to rationalize the preference of $\pi$-electron density in
C$_{60}$ to lie on 6-6 bonds, which explains the observed reactivity
patterns. They have also noted the importance of anti-ferromagnetic
spin-spin correlations and how these can be well described in terms of
the resonace of valence bond structures such as those characterizing
the Kekul\'e basis.

In recent {\em ab initio} studies, the electronic structure of
polyacene molecules composed of linearly fused benzene units has been
investigated
\cite{Hachmann-2007,Gidofalvi-2008,Rivero-2013,Mizukami-2013,Plasser-2013,Horn-2014,Small-2014}. In
particular, we and others have discussed how these systems become
polyradical in nature using multi-reference {\em ab initio}
methods. According to Hachmann {\em et al.} \cite{Hachmann-2007}, the
nature of the chemical bond in the $\pi$-space of polyacene molecules
is best understood in terms of a localized resonating valence bond
structure, rather than a more traditional delocalized-electron picture
which would give rise to metallic-like behavior. We note that even HF
predicts, in its own poor-man's approach, the polyradical character of
these systems: a RHF solution becomes unstable towards a UHF
description in which electrons localize in the $p$ orbitals of the
carbon atoms according to an anti-ferromagnetic pattern. The broken
symmetry HF description is characterized by long-range magnetic
ordering.

In light of all these previous works, we decided to try a broken
symmetry HF approach for the fullerene systems in full {\em ab initio}
calculations. It was natural to consider non-collinear type solutions
given the frustration induced by the presence of pentagon rings or,
equivalently, the curvature present in these structures. Such an
orientation can only be described in the framework of generalized HF
(GHF) solutions
\cite{Fukutome-1981,Lowdin-1992,HammesSchiffer-1993,Stuber-2003},
where each molecular orbital becomes a spinor. That is, each molecular
orbital $\phi_i (\mathbf{r})$ is expanded, in the linear-combination
of atomic orbitals (LCAO) approach, as
\begin{equation}
  \phi_i (\mathbf{r}) = \sum_\mu \left( C_{\mu \uparrow,i} \,
  |\uparrow \rangle + C_{\mu \downarrow,i} \, |\downarrow \rangle
  \right) \chi_\mu (\mathbf{r}),
\end{equation}
where $\chi_\mu (\mathbf{r})$ is an atom-centered basis function and
$C_{\mu \uparrow,i}$ and $C_{\mu \downarrow,i}$ are independent
coefficients. The use of GHF solutions in chemistry has been rather
sparse. We recently noted \cite{JimenezHoyos-2011} that GHF solutions
must be invoked in order to fix the size-consistency problem of UHF in
certain molecules, such as O$_2$. Yamaguchi and coworkers (see, {\em
  e.g.}, Refs. \cite{Yamaguchi-1999,Kawakami-2009}) have described how
non-collinear HF solutions appear naturally in frustrated systems
involving metal centers or organic radicals. Interestingly enough, in
Ref. \cite{Yamaguchi-1999} GHF-type solutions were described for small
carbon clusters. It was anticipated that a non-collinear solution
would be suppressed in C$_{60}$ due to bond alternation, in spite of
the presence of pentagon rings.

In this work we show, in full {\em ab initio} calculations, that HF
affords broken symmetry solutions of the generalized type for
fullerene molecules which are lower in energy than their RHF or UHF
counterparts. We explore the nature of these solutions in terms of the
arrangement of the localized spins in order to minimize the
frustration induced by the presence of pentagon rings. We discuss how
HF predicts these localized spins to interact via short-range
spin-spin interactions, thus giving a picture consistent with previous
model Hamiltonian results. These spin-spin interactions induce
long-range magnetic ordering in the HF solution, leading to an overall
polyradical character in the system. Lastly, we stress on the
importance to carry out a spin projection in order to remove the
unphysical effects associated with UHF or GHF solutions.

\section{Computational details}

We have performed RHF, UHF, and GHF geometry optimizations using a
modified version of the \verb+Gaussian 09+ \cite{gaussian} suite of
programs. The corresponding HF wave functions have been obtained using
a quasi-Newton optimization method \cite{JimenezHoyos-unp}. The
6-31G($d$) and the 6-311G($d$) basis sets, using Cartesian gaussian
functions, were used in the calculations. Real orbitals were used with
RHF and UHF, but complex solutions were used for GHF \footnote{Note
  that complex solutions are required in order to allow the spin
  orientations to adopt a 3D structure.}. We verified that the RHF and
UHF solutions obtained were stable (in the Slater determinant space)
and that the corresponding optimized geometries were true local
minima. Unfortunately we were not able to perform such tests in GHF
wave functions and geometries, as \verb+Gaussian 09+ currently lacks
such capabilities.

For selected systems we have computed single-point symmetry-projected
HF \cite{JimenezHoyos-2012} energies and/or natural orbital
occupations resulting from the optimized GHF solutions. We use the
acronym S-GHF to denote {\em spin} (S) projected calculations out of
GHF-type solutions. Here, the spin symmetry projection was carried out
using the ``transfer'' operators
\begin{equation}
  \hat{P}^s_{mk} = \frac{2s+1}{8\pi^2} \int \mathrm{d}\Omega \,
  D^{s\ast}_{mk} (\Omega) \, \hat{R}(\Omega),
\end{equation}
where $\Omega = (\alpha, \beta, \gamma)$ is the set of Euler angles,
$D^s_{mk} (\Omega)$ are Wigner matrices, and
\begin{equation}
  \hat{R}(\Omega) = \exp(-i\alpha\hat{S}_z) \,
  \exp(-i\beta\hat{S}_y) \, \exp(-i\gamma\hat{S}_z)
\end{equation}
is the spin rotation operator. Note that the S-GHF wavefunction is
written as a superposition of nonorthogonal Slater determinants. The
projector above was discretized using sufficient points to converge
expectation values such as $\langle \hat{S}^2 \rangle$ to high
accuracy ($10^{-8}$ or better).  Due to their expensive nature, all
the calculations in this work are in a projection-after-variation
framework \cite{JimenezHoyos-2012}, using the GHF orbitals without
further reoptimization. Such calculations were carried out using an
in-house program \cite{JimenezHoyos-2013}.

We have used the structures provided on Ref. \cite{mich} as initial
guess geometries for some of the fullerenes considered. The Schlegel
diagrams used in some of the figures were prepared with the help of
the \verb+Fullerene+ program \cite{fullerene}. The molecular structure
images shown in this paper, as well as the figure displayed in the
table of contents, were prepared using \verb+XCrySDen+
\cite{xcrysden}.

{\em Population analysis}. Expectation values of one-body operators
can be computed using the one-particle density matrix. For instance,
the expectation values of the spin operators $\hat{S}_x$, $\hat{S}_y$,
and $\hat{S}_z$ are given by
\begin{equation}
  \langle \hat{\mathbf{S}} \rangle = \frac{1}{2} \sum_{\mu \nu}
  \sum_{s s'} \boldsymbol{\sigma}_{s s'} \, S_{\mu \nu} \,
  \gamma^1_{\nu s', \mu s},
\end{equation}
where $S$ is the overlap matrix, $\boldsymbol{\sigma}$ are the Pauli
matrices, and $\gamma^1$ is the one-particle density matrix in the
atomic spin-orbital basis. The latter is related to the real-space
one-particle reduced density matrix \cite{Lowdin-1955} by
\begin{align}
  \Gamma^1 (\mathbf{r} s, \mathbf{r'} s')
  &\equiv \, N \int \mathrm{d}\mathbf{x}_2 \cdots
  \mathrm{d}\mathbf{x}_N \, \Psi^\ast (\mathbf{r} s, \mathbf{x}_2,
  \ldots, \mathbf{x}_N) \, \Psi (\mathbf{r'} s', \mathbf{x}_2, \ldots,
  \mathbf{x}_N) \nonumber \\
  &= \, \sum_{\mu \nu} \gamma^1_{\nu s', \mu s} \, \chi^\ast_{\mu}
  (\mathbf{r}) \, \chi_{\nu} (\mathbf{r'}),
\end{align}
where $N$ is the number of electrons in the system.

We define an atomic magnetic moment $\mathbf{M}_A$ using a
Mulliken-like scheme, where $S_{\mu \nu} \, \gamma^1_{\nu s', \mu s}$
is interpreted as a population matrix \cite{Mulliken-1955,Jensen}. The
magnetic moments are given by
\begin{align}
  \mathbf{M}_A
  &= \, \frac{1}{4} \sum_{\mu \in A, \nu} \sum_{s s'}
  \boldsymbol{\sigma}_{s s'} \, S_{\mu \nu} \, \gamma^1_{\nu s', \mu
    s} \nonumber \\
  &+ \, \frac{1}{4} \sum_{\nu \in A, \mu} \sum_{s s'}
  \boldsymbol{\sigma}_{s s'} \, S_{\mu \nu} \, \gamma^1_{\nu s', \mu
    s},
\end{align}
where a symmetrization was done (compared to the standard Mulliken
scheme) in order to obtain real values for $\mathbf{M}_A$. Note that,
as long as only atom-centered functions are used, the sum of the
atomic magnetic moments is equal to the corresponding spin expectation
value, i.e., $\langle \hat{S}_z \rangle = \sum_A M^z_A$.

Expectation values of two-body operators can be computed, analogously,
using the two-particle density matrix. For instance,
\begin{equation}
  \langle \hat{S}^2 \rangle = \frac{1}{2} \sum_{\mu \nu \lambda
    \kappa} \, \sum_{j=x,y,z} \, \sum_{s_1 s'_1} \sum_{s_2 s'_2}
  \sigma^j_{s_1 s'_1} \, \sigma^j_{s_2 s'_2} \, S_{\mu \nu} \,
  S_{\lambda \kappa} \, \tilde{\gamma}^2_{\nu s'_1 \kappa s'_2, \mu
    s_1 \lambda s_2},
\end{equation}
where
\begin{equation}
  \gamma^2_{\nu s'_1 \kappa s'_2, \mu s_1 \lambda s_2} =
  \tilde{\gamma}^2_{\nu s'_1 \kappa s'_2, \mu s_1 \lambda s_2} -
  \frac{1}{2} \gamma^1_{\kappa s'_2, \mu s_1} \, S^{-1}_{\nu \lambda}
  \, \delta_{s'_1 s_2}
\end{equation}
is the two-particle density matrix in the atomic spin-orbital basis
(normalized to the number of independent electron pairs). The latter
is related to the real-space two-particle reduced density matrix
\cite{Lowdin-1955} by
\begin{equation}
  \Gamma^2 (\mathbf{r}_1 s_1 \mathbf{r}_2 s_2, \mathbf{r'}_1 s'_1
  \mathbf{r'}_2 s'_2)
  = \, \sum_{\mu \nu}
  \gamma^2_{\nu s'_1 \kappa s'_2, \mu s_1 \lambda s_2} \,
  \chi^\ast_{\mu} (\mathbf{r}_1) \, \chi^\ast_{\lambda} (\mathbf{r}_2)
  \, \chi_{\nu} (\mathbf{r'}_1) \, \chi_{\kappa} (\mathbf{r'}_2).
\end{equation}

Analogous to the one-particle case, one can interpret $S_{\mu \nu} \,
S_{\lambda \kappa} \, \gamma^2_{\nu s'_1 \kappa s'_2, \mu s_1 \lambda
  s_2}$ as a two-electron population matrix. We compute atomic
spin-spin correlations $\mathbf{S}_A \cdot \mathbf{S}_B$ in the form
\begin{align}
  \mathbf{S}_A \cdot \mathbf{S}_B
  &= \, \frac{1}{4} \sum_{\mu \in A, \lambda \in B} \sum_{\nu \kappa} \,
  \sum_{j=x,y,z} \, \sum_{s_1 s'_1} \sum_{s_2 s'_2} \sigma^j_{s_1
    s'_1} \, \sigma^j_{s_2 s'_2} \, S_{\mu \nu} \, S_{\lambda \kappa}
  \, \tilde{\gamma}^2_{\nu s'_1 \kappa s'_2, \mu s_1 \lambda s_2}
  \nonumber \\
  &+ \, \frac{1}{4} \sum_{\nu \in A, \kappa \in B} \sum_{\mu \lambda} \,
  \sum_{j=x,y,z} \, \sum_{s_1 s'_1} \sum_{s_2 s'_2} \sigma^j_{s_1
    s'_1} \, \sigma^j_{s_2 s'_2} \, S_{\mu \nu} \, S_{\lambda \kappa}
  \, \tilde{\gamma}^2_{\nu s'_1 \kappa s'_2, \mu s_1 \lambda s_2},
  \label{eq:spin-spin}
\end{align}
where a symmetrization was also carried out in order to obtain real
values. Note that $\langle \hat{S}^2 \rangle = \sum_{AB} (\mathbf{S}_A
\cdot \mathbf{S}_B)$.

\section{Results and discussion}

We present HF solutions of the generalized type for C$_{84}$
($D_{6h}$), C$_{70}$ ($D_{5h}$), C$_{60}$ ($I_h$), C$_{36}$
($D_{6h}$), C$_{30}$ ($D_{5h}$), and for four isomers of C$_{20}$ (the
cage, the bowl, the ring, and a recently proposed pentagon structure
\cite{Cardenas-2012}). We note that a $D_2$ and a $D_{2d}$ isomer of
C$_{84}$ are known to be lower in energy than the $D_{6h}$ structure
considered in this work \cite{Raghavachari-1992}. The features of the
optimized geometries are described in Sec. \ref{sec:geom}. In
Sec. \ref{sec:ener} we contrast the energies obtained by GHF with
those from RHF. In Secs. \ref{sec:moment}, \ref{sec:occ}, and
\ref{sec:spin} we describe the polyradical character present in the
GHF solutions in terms of the local magnetic moments, natural
occupations, and atomic spin-spin correlation functions,
respectively. Lastly, in Sec. \ref{sec:disc} we provide a brief
discussion on the interpretation of our results. In what follows, we
provide a short review of the theoretical work on the smaller
fullerenes.

The sucessful synthesis in 1998 of a solid form of C$_{36}$
\cite{Piskoti-1998} prompted theoretical work on this fullerene
\cite{Jagadeesh-1999,Fowler-1999,Fowler-1999b,Ito-2000,Slanina-2000,Varganov-2002}. From
early on it was recognized that the $D_{6h}$ isomer possesses
significant diradical character. After some disagreement in the
geometry and even the multiplicity of the ground state, it was
recognized that correlation effects play a very significant role in
determining the correct geometry and electronic properties of this
molecule \cite{Varganov-2002}. Fewer theoretical studies have focused
on C$_{30}$ due to its triplet ground state character according to
H\"uckel theory \cite{Fan-1995} and its observed low kinetic stability
\cite{Lu-2005}. R(O)HF also predicts a triplet ground state for the
$D_{5h}$ structure \cite{Paulus-2003}.

Since the early days of fullerene science, C$_{20}$ became an
interesting subject for theoretical studies for at least two reasons:
a) in order to understand the processes that drive the formation of
C$_{60}$ \cite{Brabec-1992}, and b) to determine what the smallest
carbon cluster is for which a cage-like structure becomes the
ground-state \cite{Feyereisen-1992}. In particular, the C$_{20}$ cage
is the smallest possible fullerene that can be constructed out of only
pentagon and hexagon rings: it has 12 pentagons (as required by
Euler's theorem) and no hexagons. After the realization that different
methods predicted different stability patterns among the cage, ring,
and bowl isomers \cite{Raghavachari-1993,Grossman-1995,Taylor-1995}, a
large amount of theoretical work followed
\cite{Bylaska-1996,Wang-1996,Martin-1996,Jones-1997,Murphy-1998,Galli-1998,Sokolova-2000}. In
2000, Prinzbach {\em et al.} \cite{Prinzbach-2000} reported the
successful synthesis of the C$_{20}$ fullerene cage from its
perhydrogenated form. Later works helped to confirm the cage-like
character of the reported structure \cite{Saito-2001,Lu-2003}, or
proposed further ways to characterize it
\cite{Romero-2002,Castro-2002}. Some recent papers have established,
using high-level correlated methods, the cage and the bowl to be
nearly isoenergetic at zero temperature
\cite{Grimme-2002,An-2005}. Zhang, Sun, and Cao performed an extensive
study of the Jahn-Teller distortions in the C$_{20}$ cage
\cite{Zhang-2007}. We note the work of Heaton-Burgess and Yang
\cite{HeatonBurgess-2010}, who have associated the delocalization
error to the incorrect energetic ordering predicted by density
functional approximations. Using semi-empirical calculations, Greene
and Beran \cite{Greene-2002} proposed a mechanism for the conversion
between the ring and the bowl. The C$_{20}$ cage has also been
investigated \cite{LopezSandoval-2006,Lin-2007,Lin-2008} in the
context of the Hubbard Hamiltonian, where it has become evident that
electron correlation in the C$_{20}$ cage plays a larger role than in
its C$_{60}$ counterpart.

\subsection{Optimized geometries}
\label{sec:geom}

Despite the many works in the literature studying the lowest-energy
C$_{20}$ isomers, relatively few of them have focused on the
structural differences predicted by different {\em ab initio}
methods. In a recent work, Grimme and M\"uck-Lichtenfeld
\cite{Grimme-2002} pointed out that the RHF/6-31G($d$) structures
considered by previous researchers might prove unreliable in order to
obtain the correct energetic ordering with single-point correlated
calculations. In particular, the geometries of the bowl and the cage
were significantly changed when optimized with a correlated
method. The authors ended up advocating the use of MP2 geometries
obtained with a triple-zeta basis set.

\begin{figure}
  \includegraphics[width=6cm]{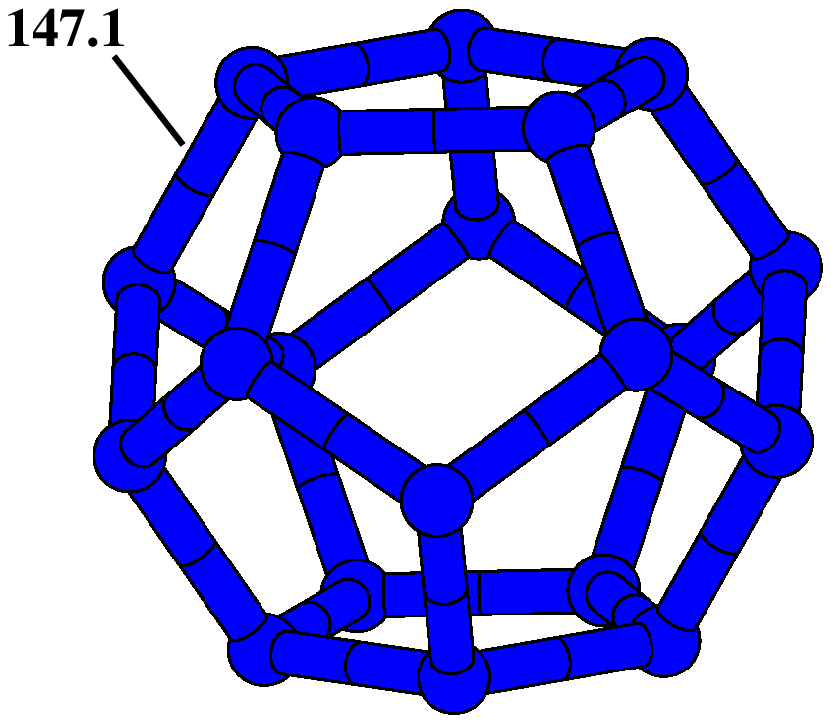}
  \hspace{0.5cm} \includegraphics[width=6cm]{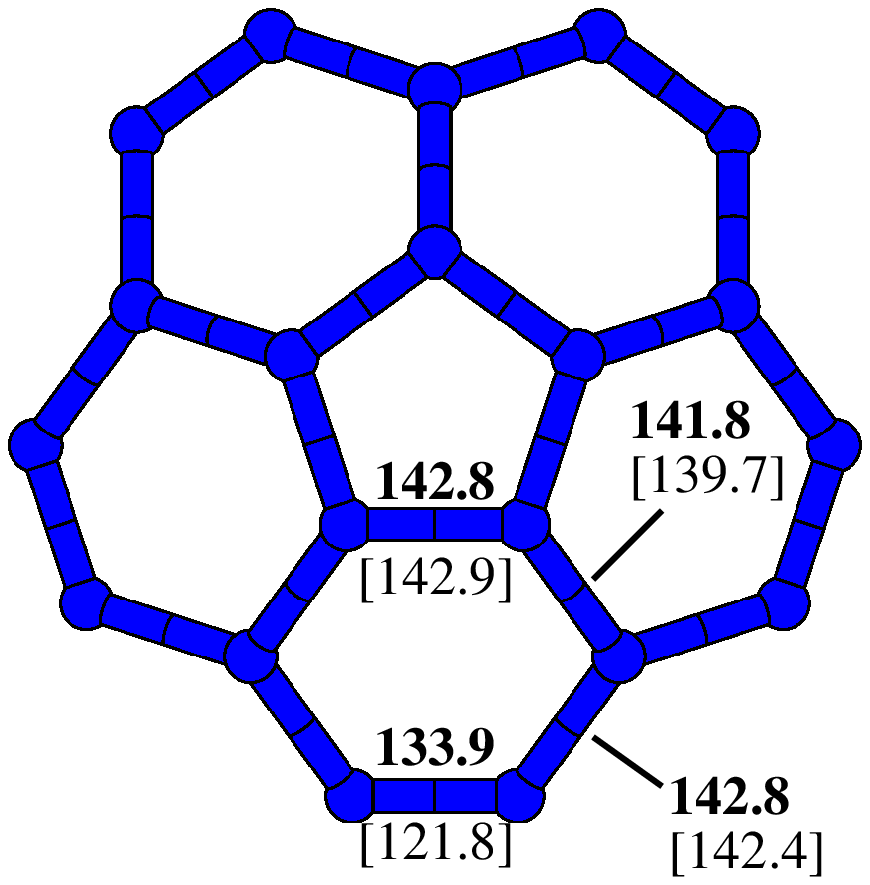}
  \\
  \includegraphics[width=6cm]{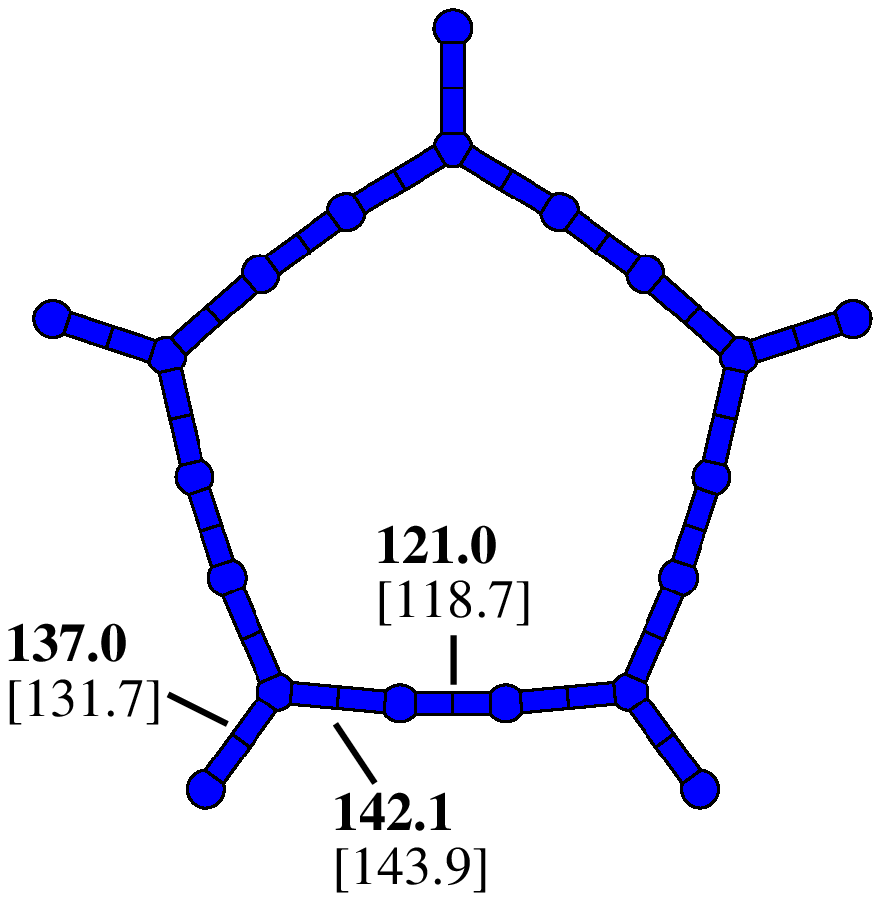}
  \hspace{0.5cm} \includegraphics[width=6cm]{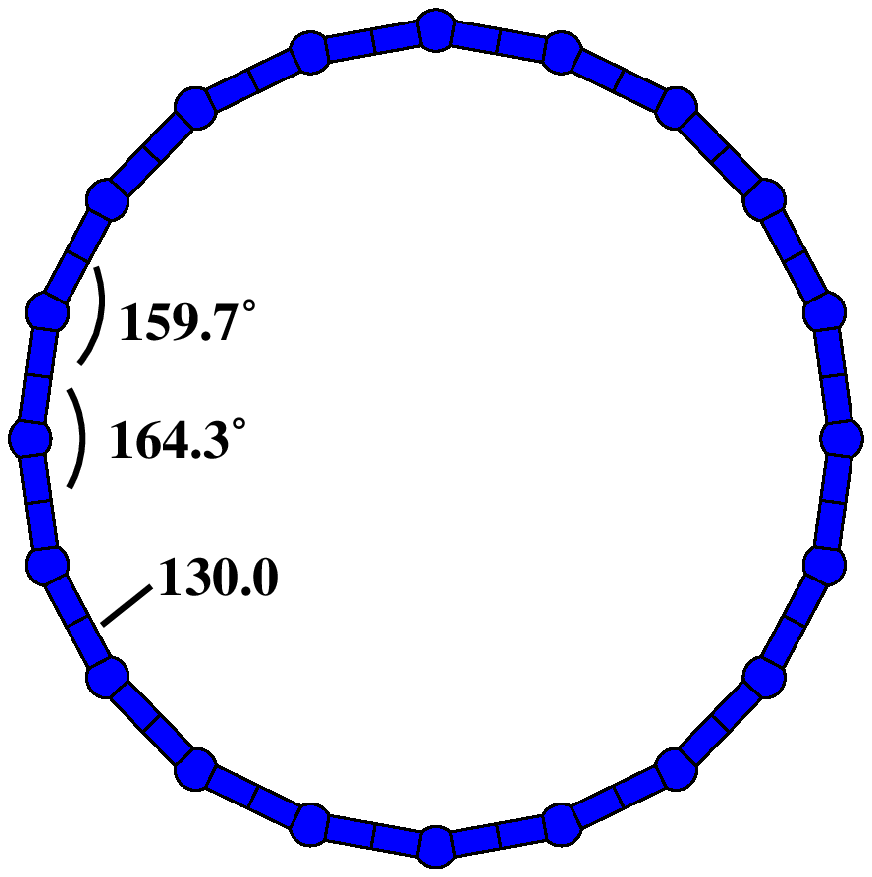}
  \caption{Structures of the four different C$_{20}$ isomers
    considered in this work, with bond lengths given in pm optimized
    at the GHF/6-31G($d$) level. The top left panel corresponds to the
    fullerene cage; GHF predicts a perfect dodecahedron with $I_h$
    symmetry. In the top right panel we show the bowl isomer
    ($C_{5v}$). The bottom left panel depicts the recently proposed
    \cite{Cardenas-2012} pentagon-like isomer ($D_{5h}$), while the
    bottom right panel shows the C$_{20}$ ring isomer ($D_{10h}$). The
    latter is slightly deviated from a $D_{20h}$ structure, as can be
    seen from the two different angles provided; all C-C distances are
    identical. For the bowl and pentagon isomers RHF/6-31G($d$) bond
    lengths are also provided in brackets for comparison
    purposes. \label{fig:c20-geom}}
\end{figure}

We show in Fig. \ref{fig:c20-geom} the optimized GHF geometries for
each of the four C$_{20}$ isomers considered in this work. At the RHF
level, the dodecahedral fullerene cage undergoes a first-order
Jahn-Teller distortion yielding an optimized structure with reduced
symmetry. Different authors have obtained different point groups for
the optimized structure; we do not discard the possibility of there
being several low-lying minima even at the RHF/6-31G($d$) level. Our
RHF/6-31G($d$) calculations yielded a $C_3$ structure, with C-C bond
distances in the range between $136.1$~pm and $150.5$~pm. On the other
hand, GHF predicts a perfect dodecahedron ($I_h$ symmetry), with all
C-C bond distances equal to $147.1$~pm. Re-optimizing the structures
with the 6-311G($d$) basis yielded only minimal changes, suggesting
that they are well converged at the HF level. We note that Lin {\em et
  al.} \cite{Lin-2007} had already suggested that if electron
correlation was strong enough in this system, the molecule would be
stable against Jahn-Teller distortions. The GHF solution displays such
character, as will be discussed in detail below.

The C$_{20}$ bowl is predicted to have a $C_{5v}$ configuration at the
RHF level of theory, deviating from a planar structure. A similar
geometry is also obtained with GHF. Nevertheless, RHF/6-31G($d$)
predicts a very short bond-length ($121.8$~pm) between the carbon
atoms located in the edges of the bowl. GHF, on the contrary, yields a
double-bond character between such atoms, with a much longer optimized
bond distance ($133.9$~pm). Switching to the 6-311G($d$) basis, the
C-C bonds at the edges are shortened to $121.3$~pm and $133.3$~pm with
RHF and GHF, respectively.

For the C$_{20}$ ring, RHF/6-31G($d$) predicts a minimum with
$C_{10h}$ structure with alternating short ($119.6$~pm) and long
($138.1$~pm) bonds. The bonds are shortened to $119.1$~pm and
$137.7$~pm upon enlarging the basis to 6-311G($d$). GHF/6-31G($d$), on
the other hand, predicts a minimum with $D_{10h}$ symmetry only
slightly distorted from the ideal $D_{20h}$ symmetry possible. In
fact, all carbon-carbon bond lengths are predicted to be
equivalent. Ten of the carbon atoms lie in a circumference with a
radius of $416.6$~pm, while the other ten atoms lie in a slightly
smaller circle with radius of $414.0$~pm. With the 6-311G($d$) basis,
GHF predicts a structure with perfect $D_{20h}$ symmetry and C-C bond
distances of $129.5$~pm.

For the C$_{20}$ pentagon, both RHF and GHF predict, with the
6-31G($d$) basis, the highly symmetric $D_{5h}$ structure to be a true
minimum. The two methods yield again a very large difference in the
bond distance between the carbon atoms located in the corner of the
pentagon (see Fig. \ref{fig:c20-geom}). Both RHF and GHF predict, with
the 6-311G($d$) basis, slightly shorter bonds for the edge acetylenic
and the corner C-C units.

\begin{figure}
  \includegraphics[width=5.4cm]{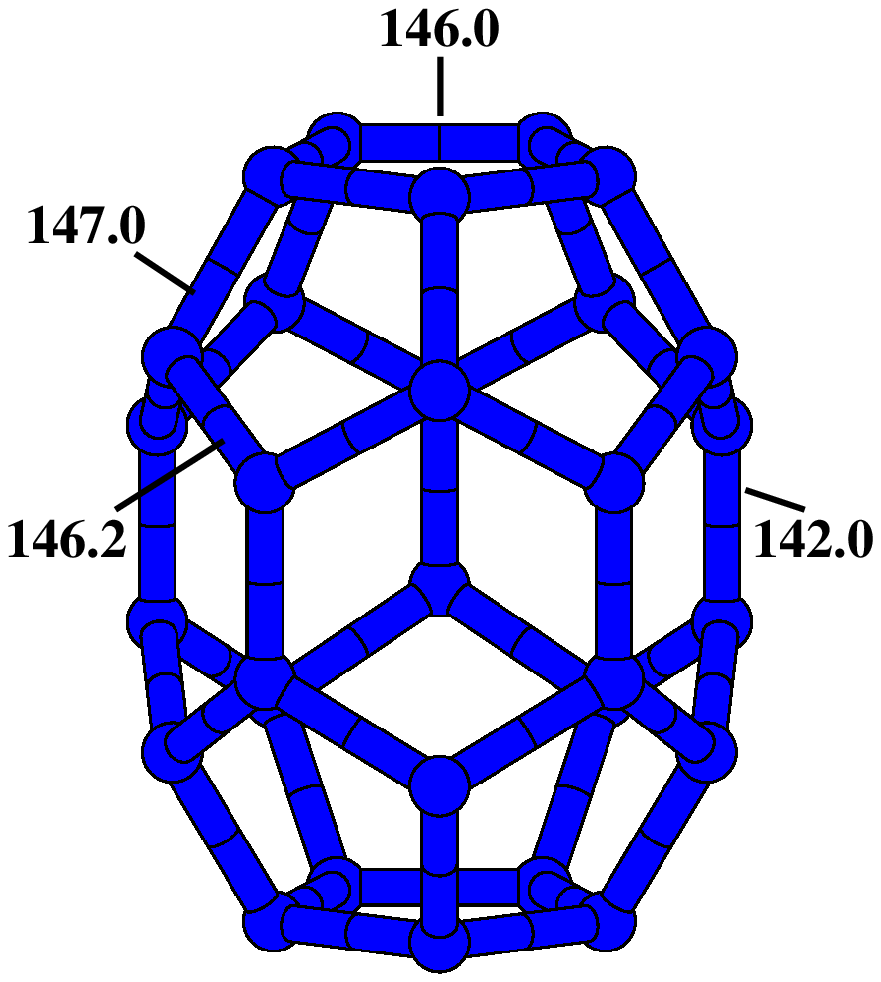}
  \hspace{0.5cm} \includegraphics[width=5.4cm]{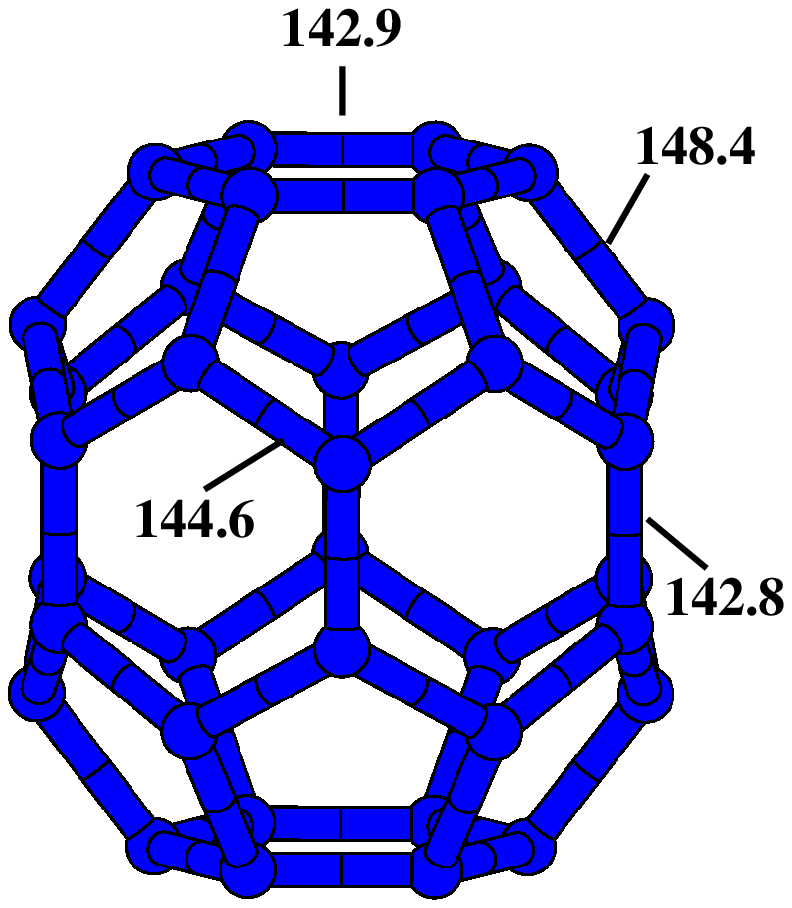}
  \caption{Structure of the C$_{30}$ ($D_{5h}$, left) and C$_{36}$
    ($D_{6h}$, right) fullerene cages considered in this work. The
    GHF/6-31G($d$) optimized geometries do not display any Jahn-Teller
    distortions to lower-symmetry structures. \label{fig:cage-geom}}
\end{figure}

We show in Fig. \ref{fig:cage-geom} the optimized geometries predicted
at the GHF/6-31G($d$) level for C$_{30}$ and C$_{36}$. It is
noteworthy that, in both cases, GHF yields the fully symmetric
geometry as a true minimum, whereas RHF (and even UHF) undergoes some
form of Jahn-Teller distortion. The most notable difference between
the GHF and RHF optimized structures for C$_{30}$ occurs in the
capping pentagons. All C-C bonds are equivalent in GHF and possess a
single-bond like character ($146.0$~pm), whereas RHF yields bond
lengths with double- ($135.1$~pm) and single-bond character
($147.6$~pm). For C$_{36}$, both RHF and GHF predict structures with a
$C_6$ axis of symmetry, though RHF only converges to a $C_{6v}$
structure as opposed to the $D_{6h}$ symmetry of the GHF
structure. The carbon-carbon bonds in the capping hexagon are
predicted to be significantly shorter by RHF ($138.9$~pm) than GHF
($142.9$~pm). We note that for both C$_{30}$ and C$_{36}$ the
difference between the GHF/6-31G($d$) and GHF/6-311G($d$) structures
is minimal.

For buckminsterfullerene (C$_{60}$) (not shown), both RHF/6-31G($d$)
and GHF/6-31G($d$) predict an icosahedral geometry ($I_h$). The bond
lengths predicted are significantly different, however. RHF predicts a
short bond-length of $137.3$~pm for hexagon-hexagon edges and a long
bond-length of $144.9$~pm for hexagon-pentagon edges. GHF, on the
other hand, predicts longer carbon-carbon bonds of $140.7$~pm and
$145.2$~pm, respectively, in good agreement with experimental
gas-phase bond lengths \cite{Hedberg-1991}. These geometrical features
are again preserved upon enlarging the basis to 6-311G($d$). It is
interesting to notice that the GHF predicted bond lengths are similar
to those reported by H\"aser {\em et al.} at the MP2/TZP level
\cite{Haser-1991}.

For C$_{70}$ and C$_{84}$ (not shown), there are again significant
structural differences between the RHF/6-31G($d$) and the
GHF/6-31G($d$) optimized geometries. For instance, RHF yields bond
lengths as short as $136.2$~pm and as long as $147.2$~pm for
C$_{70}$. GHF, on the other hand, yields bond lengths in the interval
$140.1$--$146.0$~pm, displaying a tendency towards bond length
equalization. For C$_{84}$, similar results are obtained: the mean
bond lengths obtained with RHF and GHF are $142.3$ and $143.5$~pm,
respectively, while the standard deviations are $3.5$ and $2.1$~pm.

We conclude that the HF geometries are, to a large extent, well
converged with the 6-31G($d$) basis. The inclusion of spin-spin
correlation effects (see below), even if only approximately at the GHF
level, results in significant structural changes. Our GHF optimized
geometries are still significantly different than the MP2/TZV2d1f ones
reported in Ref. \cite{Grimme-2002} for the cage, bowl, and ring
isomers of C$_{20}$. Given the large static correlation present in
these carbon clusters (see below), we find it more appropriate to
optimize the geometries with either a multi-reference method or a
broken-symmetry GHF-type independent-particle model.

\subsection{Energetics}
\label{sec:ener}

The energetic ordering between the ring, bowl, and cage isomers of
C$_{20}$ has been the subject of numerous previous works as discussed
above. We present in Table \ref{tab:c20-ener} the total energies
predicted with RHF, GHF, and S-GHF with a 6-31G($d$) basis. It becomes
immediately apparent that all methods agree in placing the pentagon
significantly higher in energy than all other isomers.

\begin{table}
  \caption{Total energies (in hartree, $+756$) predicted for the cage,
    bowl, ring, and pentagon isomers of C$_{20}$ using the 6-31G($d$)
    basis set. Relative energies (in eV) with respect to the ring
    isomer are shown in parentheses. \label{tab:c20-ener}}
    {\small
    \begin{tabular}{l l r@{ }r r@{ }r r@{ }r r@{ }r}
      \hline \hline
      method & geom &
        \multicolumn{2}{r}{cage} &
        \multicolumn{2}{r}{bowl} &
        \multicolumn{2}{r}{ring} &
        \multicolumn{2}{r}{pentagon} \\[2pt]
      \hline
      RHF           & opt. &
        -0.556\,79 & (3.61) & -0.650\,92 & (1.05) &
        -0.689\,48 & (0.00) & -0.389\,42 & (8.17) \\[2pt]
      GHF           & RHF  &
        -0.698\,56 & (2.10) & -0.791\,60 & (-0.43) &
        -0.775\,84 & (0.00) & -0.417\,77 & (9.74) \\[2pt]
      GHF           & opt. &
        -0.763\,79 & (3.12) & -0.871\,28 & (0.19) &
        -0.878\,31 & (0.00) & -0.430\,89 & (12.17) \\[6pt]
      S-GHF ($s=0$) & GHF  &
        -0.854\,09 & (3.43) & -0.979\,49 & (0.02) &
        -0.980\,07 & (0.00) & -0.566\,44 & (11.26) \\[2pt]
      S-GHF ($s=1$) & GHF  &
        -0.826\,05 & & -0.946\,74 & &
        -0.951\,85 & & -0.491\,61 & \\[2pt]
      S-GHF ($s=2$) & GHF  &
        -0.769\,24 & & -0.903\,28 & &
        -0.895\,30 & & -0.410\,60 & \\[2pt]
      \hline \hline
    \end{tabular}}
\end{table}

RHF predicts the ring to be the most stable isomer, with the cage
being $3.61$~eV higher in energy. Using the RHF geometries, a huge
energetic improvement is observed for all isomers upon allowing for
symmetry breaking. GHF predicts the ring to be the most stable isomer
using self-consistent geometries, though it is almost isoenergetic
with the bowl. The cage is $3.12$~eV higher in energy than the
ring. We note that previous studies have predicted, using highly
correlated approaches, the bowl and cage to be nearly isoenergetic
with the ring being somewhat higher in energy
\cite{Grimme-2002,An-2005}. It would be interesting to repeat some of
these calculations with the GHF geometries reported in this work. All
GHF solutions display significant spin contamination, with $\langle
\hat{S}^2 \rangle$ equaling $6.42$, $7.70$, $7.23$, and $5.14$
a.u. for the cage, bowl, ring and pentagon isomers, respectively, at
the GHF optimized geometries.

The energetic improvement afforded by S-GHF ($s=0$) over GHF is
significant in all cases, consistent with the large spin contamination
observed in the GHF solutions. S-GHF predicts the triplet states of
the cage, bowl, and ring isomers to lie $\approx 30$~mhartree higher
in energy than their singlet counterparts. The singlet-triplet gaps
predicted by S-GHF are most likely overestimated, as the optimization
of the wave function is naturally biased towards the ground
state. Nonetheless, S-GHF clearly points towards a singlet ground
state for all isomers. The quintet states are predicted to be
significantly higher in energy in all cases.

In Table \ref{tab:large} we present a summary of the total energies
and $\langle \hat{S}^2 \rangle$ predicted by RHF, UHF, and GHF for the
larger fullerene systems. Several features are immediately apparent
from the Table. The correlation energy (here defined as the difference
with respect to the RHF energy) that GHF yields at the optimized
geometries is huge in all cases, ranging from $\approx 160$~mhartree
for C$_{60}$ to $\approx 280$~mhartree for both C$_{30}$ and
C$_{36}$. Note that most of the energetic improvement is already
recovered at the UHF level in C$_{30}$ and C$_{36}$, but not in
C$_{60}$, C$_{70}$, or C$_{84}$, where GHF still provides a
significant improvement over UHF ($50$--$60$~mhartree). This is also
reflected in the large difference in the spin contamination between
UHF and GHF for the larger fullerenes. We point out that GHF should be
preferred over UHF even for the smaller isomers in light of the highly
symmetric structures that it affords and its ability to minimize spin
frustration effects. As observed in the case of the C$_{20}$ isomers,
the energetic improvement brought upon by the geometry relaxation at
the GHF level is significant. In other words, the difference between
GHF//RHF and GHF//GHF energies is considerable in all cases. This is
consistent with the large structural differences observed between RHF
and GHF optimized geometries.

\begin{table}
  \caption{Total energies (in hartree), molecular symmetries, and
    $\langle \hat{S}^2 \rangle$ (in a.u.) as predicted by RHF, UHF,
    and GHF for the C$_{30}$, C$_{36}$, C$_{60}$, C$_{70}$, and
    C$_{84}$ fullerene cages. The 6-31G($d$) basis was used in the
    calculations. \label{tab:large}}
    {\small
    \begin{tabular}{l r@{ }r@{ }r | r@{ }r@{ }r | r@{ }r@{ }r | r}
      \hline \hline
      &
        \multicolumn{3}{r}{RHF//RHF} &
        \multicolumn{3}{r}{UHF//UHF} &
        \multicolumn{3}{r}{GHF//GHF} &
        GHF//RHF
        \\[4pt] \cline{2-4} \cline{5-7} \cline{8-10} \cline{11-11}
      \\[-10pt]
      mol &
      sym & $\langle \hat{S}^2 \rangle$ & $E$ &
      sym & $\langle \hat{S}^2 \rangle$ & $E$ &
      sym & $\langle \hat{S}^2 \rangle$ & $E$ &
      $E$ \\[2pt]
      \hline
      C$_{30}$ &
        $C_s$    &  0.00 & -1135.187\,47 &
        $C_2$    &  7.66 & -1135.467\,13 &
        $D_{5h}$ &  8.19 & -1135.468\,93 &
                           -1135.375\,30 \\[2pt]
      C$_{36}$ &
        $C_{6v}$ &  0.00 & -1362.540\,56 &
        $D_{3h}$ &  7.68 & -1362.818\,18 &
        $D_{6h}$ &  8.12 & -1362.826\,72 &
                           -1362.781\,51 \\[2pt]
      C$_{60}$ &
        $I_h$    &  0.00 & -2271.830\,39 &
        $C_i$    &  7.52 & -2271.925\,54 &
        $I_h$    & 10.26 & -2271.991\,34 &
                           -2271.964\,82 \\[2pt]
      C$_{70}$ &
        $D_{5h}$ &  0.00 & -2650.565\,86 &
        $C_s$    &  9.40 & -2650.714\,84 &
        $D_{5h}$ & 12.09 & -2650.769\,31 \\[2pt]
      C$_{84}$ &
        $D_{6h}$ &  0.00 & -3180.806\,80 &
        $D_{3d}$ & 11.62 & -3181.007\,10 &
        $D_{6h}$ & 13.99 & -3181.062\,67 \\[2pt]
      \hline \hline
    \end{tabular}}
\end{table}

Table \ref{tab:large2} displays the total energies predicted by S-GHF
for the C$_{30}$ and C$_{36}$ fullerene cages, using the smaller 6-31G
basis at the optimized GHF/6-31G($d$) geometries. The singlet state is
the lowest in energy, with the triplet state lying $24$ and
$31$~mhartree above for C$_{30}$ and C$_{36}$, respectively. S-GHF
thus agrees with multireference treatments for C$_{36}$ in placing the
singlet as the ground state \cite{Varganov-2002}; the triplet ground
state predicted by ROHF for C$_{30}$ \cite{Paulus-2003} is most likely
incorrect.

\begin{table}
  \caption{Total energies (in hartree) predicted by S-GHF for the
    C$_{30}$ and C$_{36}$ fullerene cages. The calculations were
    carried out with the 6-31G basis at the GHF/6-31G($d$) optimized
    geometries. \label{tab:large2}}
    {\small
    \begin{tabular}{l r r}
      \hline \hline
      method & C$_{30}$ & C$_{36}$ \\[2pt]
      \hline
      GHF
        & -1135.022\,96
        & -1362.302\,35 \\[2pt]
      S-GHF ($s=0$)
        & -1135.129\,03
        & -1362.434\,33 \\[2pt]
      S-GHF ($s=1$)
        & -1135.105\,25
        & -1362.403\,55 \\[2pt]
      S-GHF ($s=2$)
        & -1135.057\,66
        & -1362.346\,75 \\[2pt]
      \hline \hline
    \end{tabular}}
\end{table}

\subsection{Atomic magnetic moments}
\label{sec:moment}

At this point, it is interesting to discuss the nature of the GHF
solutions obtained for the various isomers. In particular, we shall
inverstigate whether GHF predicts some type of torsional spin-density
wave \cite{Fukutome-1981} being developed in the molecule. We note
that the nature of the GHF solution for the C$_{20}$ ring isomer (not
shown) is trully collinear, anti-ferromagnetic in character,
coinciding with a UHF-type description. In other words, an axial
spin-density wave, with alternating up- and down-spins, develops along
the ring.

\begin{figure}
  \includegraphics[width=6cm]{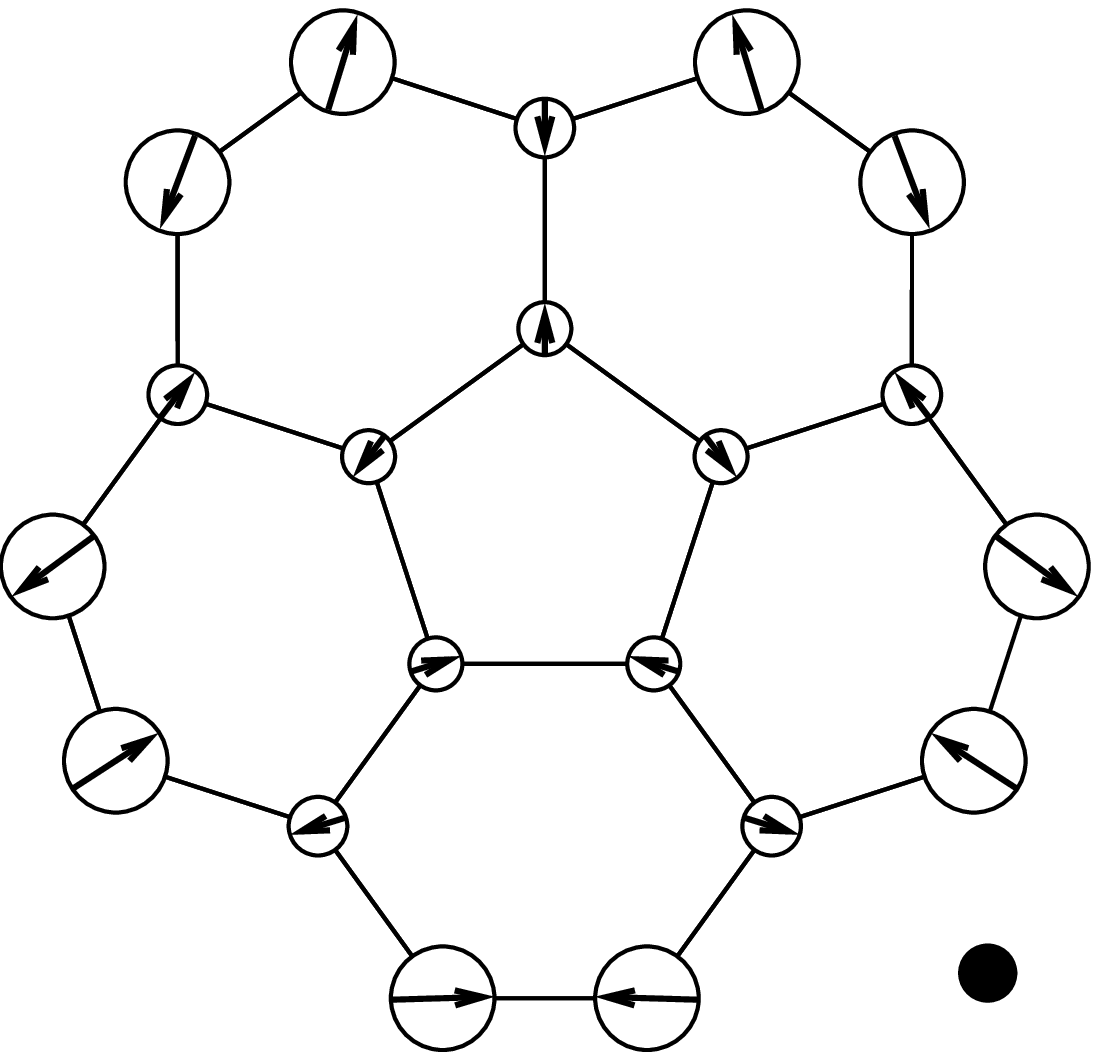}
  \hspace{0.5cm} \includegraphics[width=6cm]{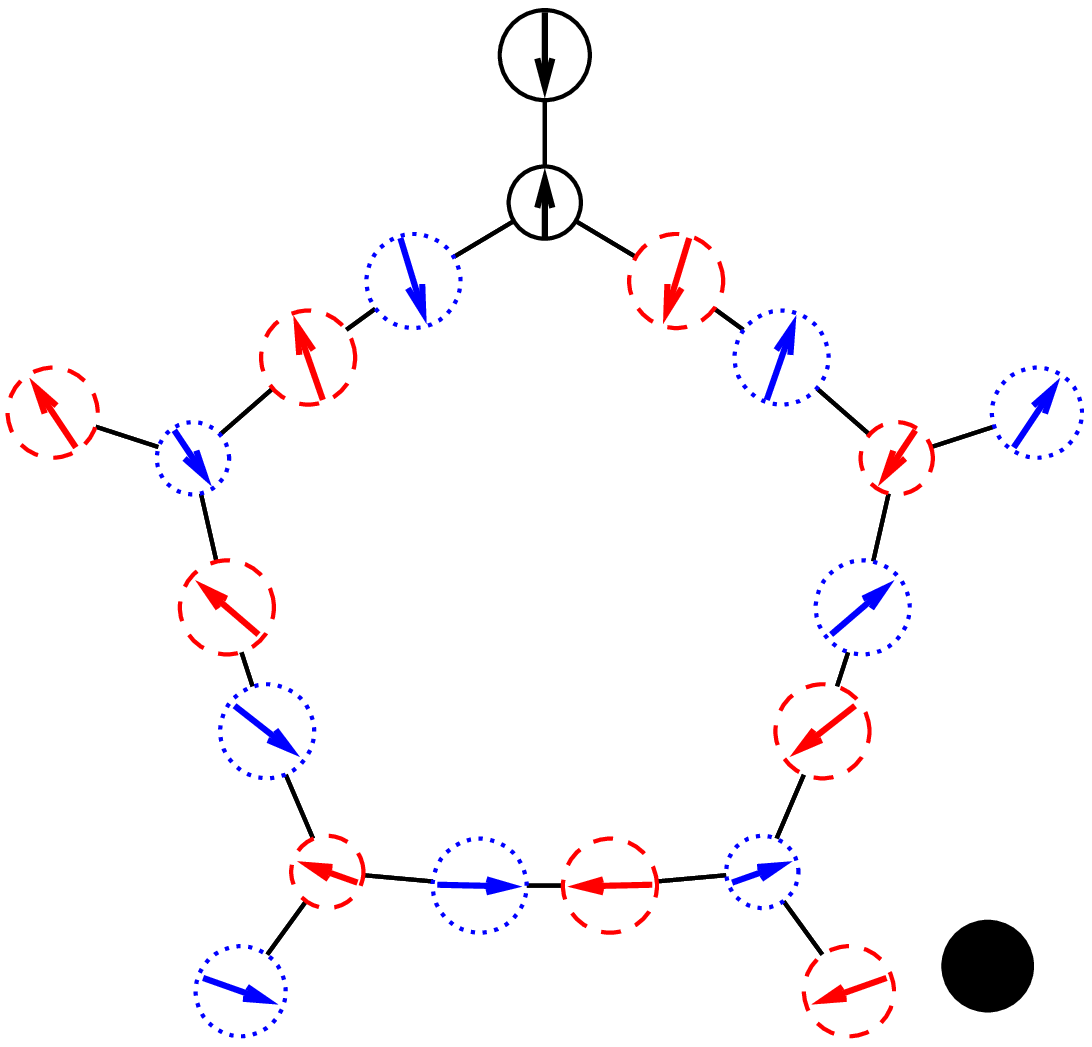}
  \caption{Atomic magnetic moments obtained from GHF/6-31G($d$)
    calculations on the C$_{20}$ bowl (left) and pentagon (right)
    isomers, at the optimized geometries. The {\em radius} of the
    circle is proportional to $\mathbf{M}_A$; the filled black circle
    to the bottom-right sets the scale ($\mathbf{M}_A = 0.5$). In each
    center, the orientation of the $xy$-projection of $\mathbf{M}_A$
    is depicted as an arrow; if the $z$-projection is positive
    (negative), the circle is displayed as dashed-red (dotted-blue). A
    solid black circle was used to indicate that there is no
    $z$-projection: the atomic moment is fully oriented in the $xy$
    plane. Note that the absolute orientation of the atomic moments is
    unimportant (in non-relativistic calculations); it is only the
    relative orientation that determines the
    physics. \label{fig:c20-ghf-struc}}
\end{figure}

Both the bowl and the pentagon C$_{20}$ isomers display a
non-collinear structure, as depicted in Fig. \ref{fig:c20-ghf-struc}.
In both isomers a torsional wave develops in order to minimize spin
frustrations, even if the nearest-neighbor interactions remain mostly
anti-ferromagnetic in character. Interestingly, all the atomic moments
in the bowl isomer are coplanar, with an enhanced spin density in the
outermost carbon atoms. For the pentagon isomer, each acetylene-like
unit along the edges has spins with a perfect anti-ferromagnetic
alignment. Although the magnitudes of the magnetic moments displayed
cannot be given a meaningful physical interpretation due to the known
problems associated with the Mulliken population analysis
\cite{Jensen}, it is interesting that the magnetic moments in the
pentagon isomer are $\approx 1/2$ in magnitude, the spin of a single
electron. This is what one would expect if there was a single unpaired
electron in a $p$ orbital of each carbon atom.

We show in Fig. \ref{fig:c20-ghf-struc} the atomic magnetic moments
obtained from the GHF solution for the C$_{20}$ cage at its optimized
geometry. Our result is similar to that presented in
Ref. \cite{LopezSandoval-2006} from the non-collinear HF solution of
the C$_{20}$ cage in a Hubbard Hamiltonian. Here, only the relative
orientation of the moments is relevant. Each moment is oriented at an
angle of $\approx 138^\circ$ with respect to its nearest neighbors,
slightly deviating from the $4\pi/5$ angle that would yield coplanar
spins in each pentagon. The structure thus obtained is the one that
minimizes the frustration to the largest extent. Note that all moments
have exactly the same magnitude, $\approx 1/2$, which provides a
rationalization for the perfectly symmetric optimized geometry
obtained from GHF.

\begin{figure}
  \includegraphics[width=6cm]{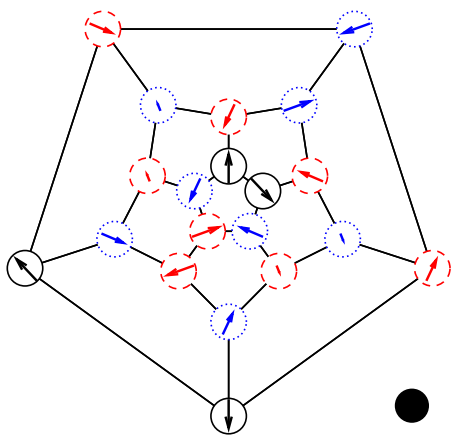}
  \caption{Same as Fig. \ref{fig:c20-ghf-struc}, but for the C$_{20}$
    fullerene cage, displayed as a Schlegel
    diagram. \label{fig:c20-ghf-struc2}}
\end{figure}

\begin{figure}
  \includegraphics[width=7cm]{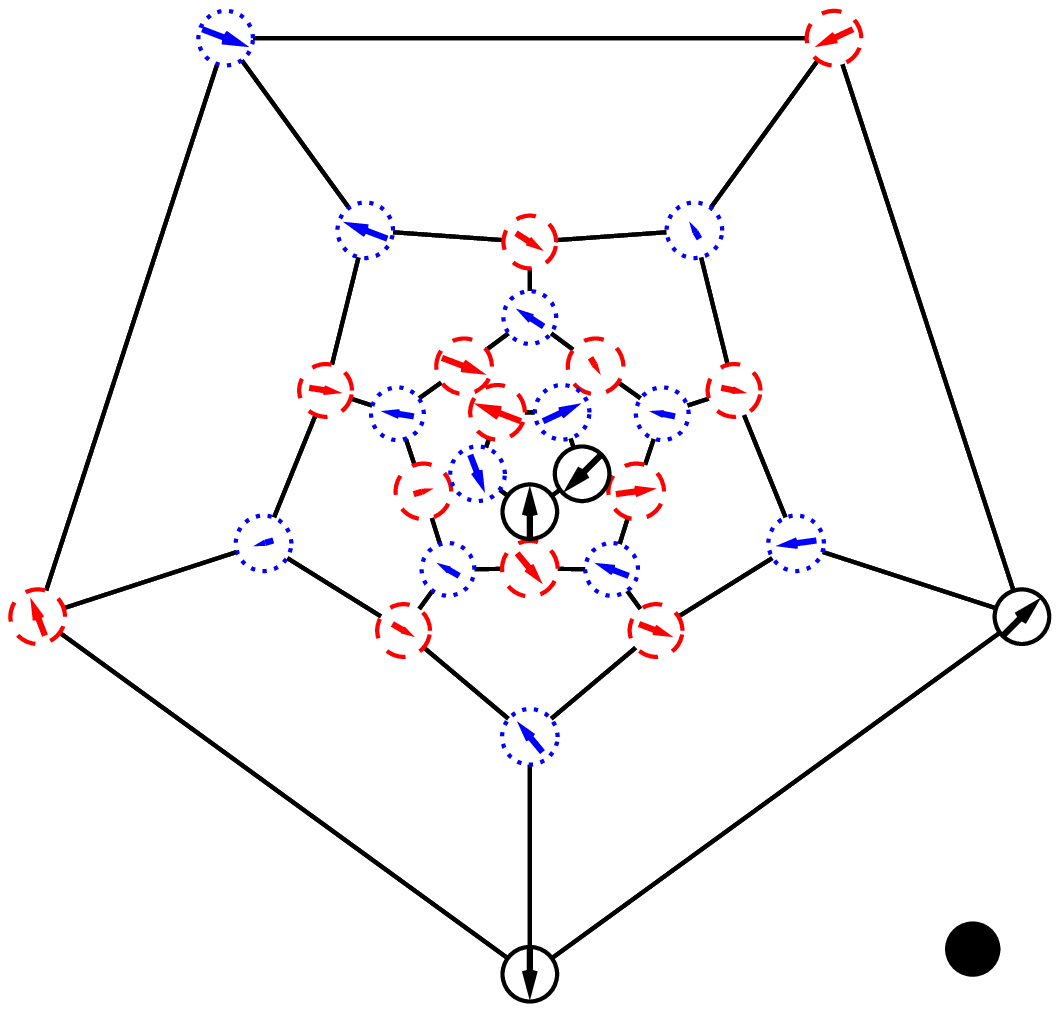}
  \hspace{0.5cm} \includegraphics[width=7cm]{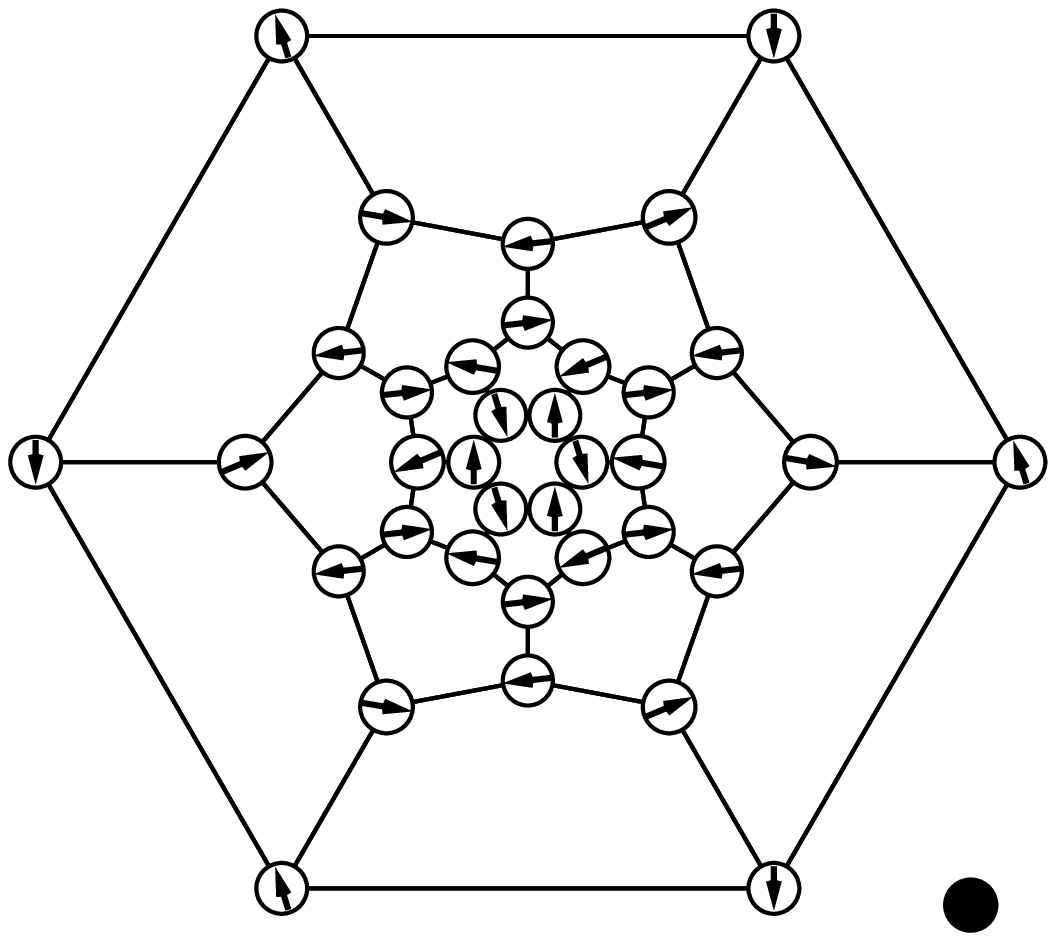}
  \caption{Same as Fig. \ref{fig:c20-ghf-struc}, but for the C$_{30}$
    (left) and C$_{36}$ (right) fullerene cages, displayed as Schlegel
    diagrams. \label{fig:large-ghf-struc}}
\end{figure}

\begin{figure}
  \includegraphics[width=7cm]{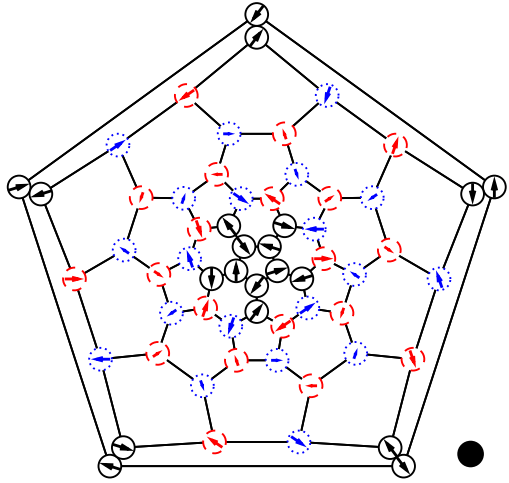}
  \caption{Same as Fig. \ref{fig:c20-ghf-struc}, but for the C$_{60}$
    fullerene cages displayed as a Schlegel
    diagram. \label{fig:large-ghf-struc2}}
\end{figure}

We present, in Figs. \ref{fig:large-ghf-struc} and
\ref{fig:large-ghf-struc2}, the atomic magnetic moments developed in
GHF/6-31G($d$) calculations of C$_{30}$, C$_{36}$, and C$_{60}$. As
expected from the fact that the GHF solution does not coincide with
UHF, all systems develop torsional spin density waves. The magnitude
of the atomic magnetic moments is again $\approx 1/2$, suggesting the
presence of a single unpaired electron.

In C$_{30}$, each atomic moment in the capping pentagon (innermost and
outermost pentagons) is oriented at an angle of $\approx 136^\circ$
with respect to its nearest neighbors, and at an angle of $\approx
70^\circ$ with respect to its next-nearest neighbors, thus minimizing
spin frustration. In the hexagonal walls, we observe perfect
anti-ferromagnetic interactions between those carbon atoms related by
the mirror plane perpendicular to the $C_5$ axis of the molecule. In
C$_{36}$, all the atomic magnetic moments lie in the same plane. In
the capping hexagon, each moment is oriented at an angle of $\approx
165^\circ$ with respect to its nearest neighbors. The moment is
parallel with respect to its next-nearest neighbors. Once again, we
find exact anti-ferromagnetic pairing between those carbon atoms in
the hexagonal walls related by the mirror plane perpendicular to the
$C_6$ molecular axis.

For C$_{60}$, the spin arrangement coincides with that described by
Coffey and Trugman \cite{Coffey-1992} from the classical Heisenberg
solution. This spin arrangement was also observed in the Hubbard
solution for sufficiently large $U/t$ (see, {\em e.g.},
Refs. \cite{Bergomi-1993,Ojeda-1999}). In particular, all spins in a
given pentagon are coplanar, but spins in neighbor pentagons lie in a
different plane. The normals to the spin planes in the pentagons are
related in a non-trivial way. Perfect anti-ferromagnetic alignment
between the carbon atoms defining the 6-6 bonds (hexagon-hexagon
edges) is observed.

\begin{figure}
  \includegraphics[width=7cm]{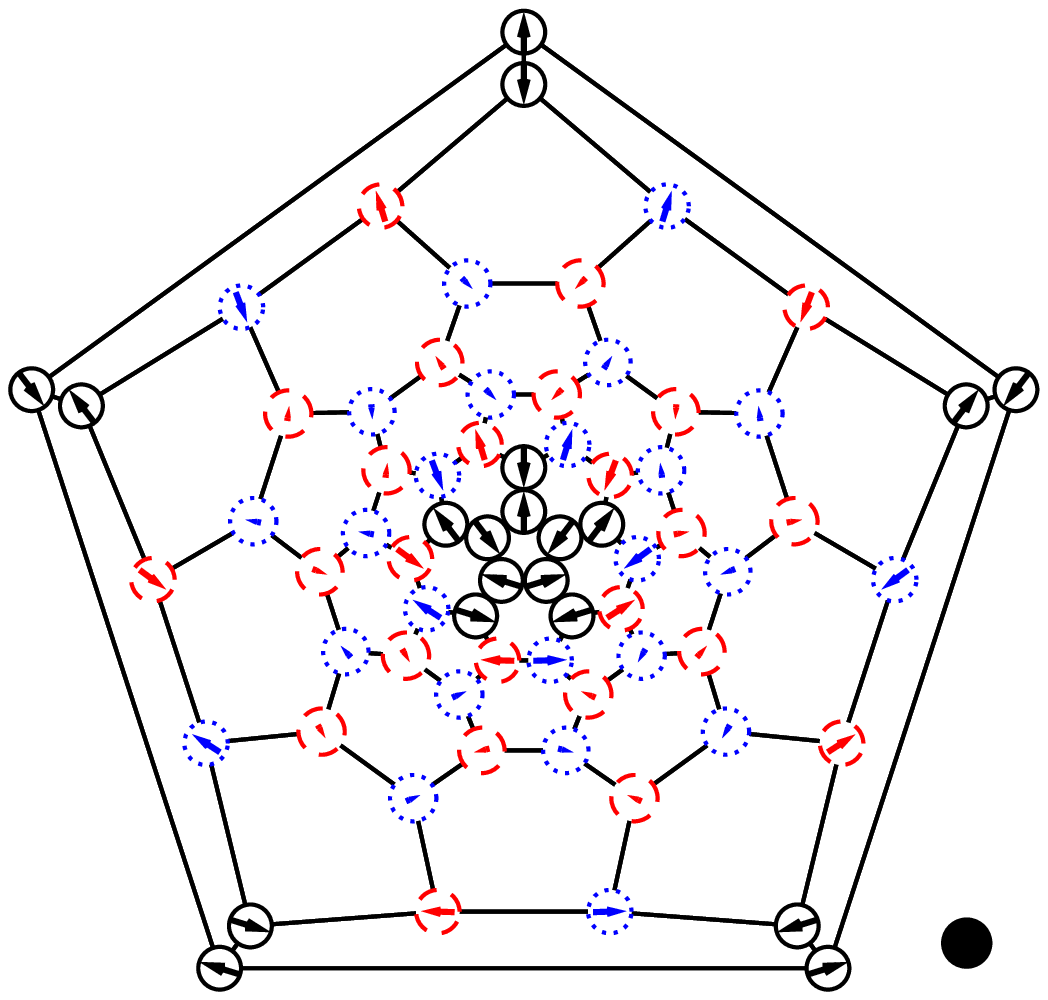}
  \hspace{0.5cm} \includegraphics[width=7cm]{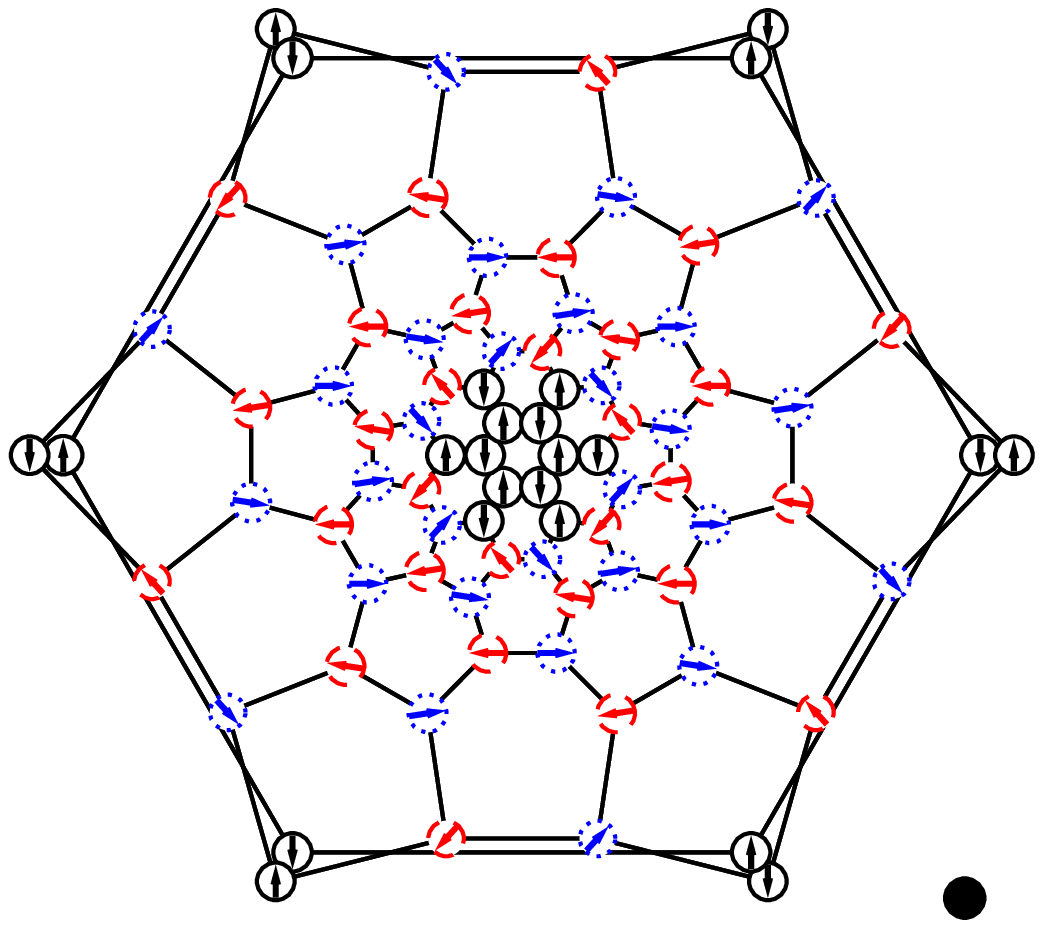}
  \caption{Same as Fig. \ref{fig:c20-ghf-struc}, but for the C$_{70}$
    (left) and C$_{84}$ (right) fullerene cages, displayed as Schlegel
    diagrams. \label{fig:large-ghf-struc3}}
\end{figure}

We show, in Fig. \ref{fig:large-ghf-struc3}, the atomic magnetic
moments obtained in GHF/6-31G($d$) calculations in C$_{70}$ and
C$_{84}$. In C$_{70}$, the pentagons in the poles of the molecular
structure have coplanar spins. The same is not true about other
pentagons, which slightly deviate form this structure. The 5-6 bonds
involving the carbon atoms in the poles display perfect
anti-ferromagnetic alignment. The moments in the 6-6 bonds in the
equatorial belt make an angle of $\approx 170^\circ$. In C$_{84}$, a
perfect N\'eel type alignment is observed in the hexagons defining the
poles. The spins in the equatorial belt also display perfect
anti-ferromagnetic alignment.

\subsection{Natural occupations}
\label{sec:occ}

One way to characterize the degree to which static correlation plays a
role in a given system is to consider the natural occupation
profile. If a (singlet) system is weakly correlated, all occupations
are expected to remain close to zero or two. Conversely, occupation
numbers near one signal the failure of (restricted) HF to become an
accurate zero-th order wave function: several determinants contribute
nearly equally to the exact wave function expansion.

\begin{figure}
  \includegraphics[width=13.8cm]{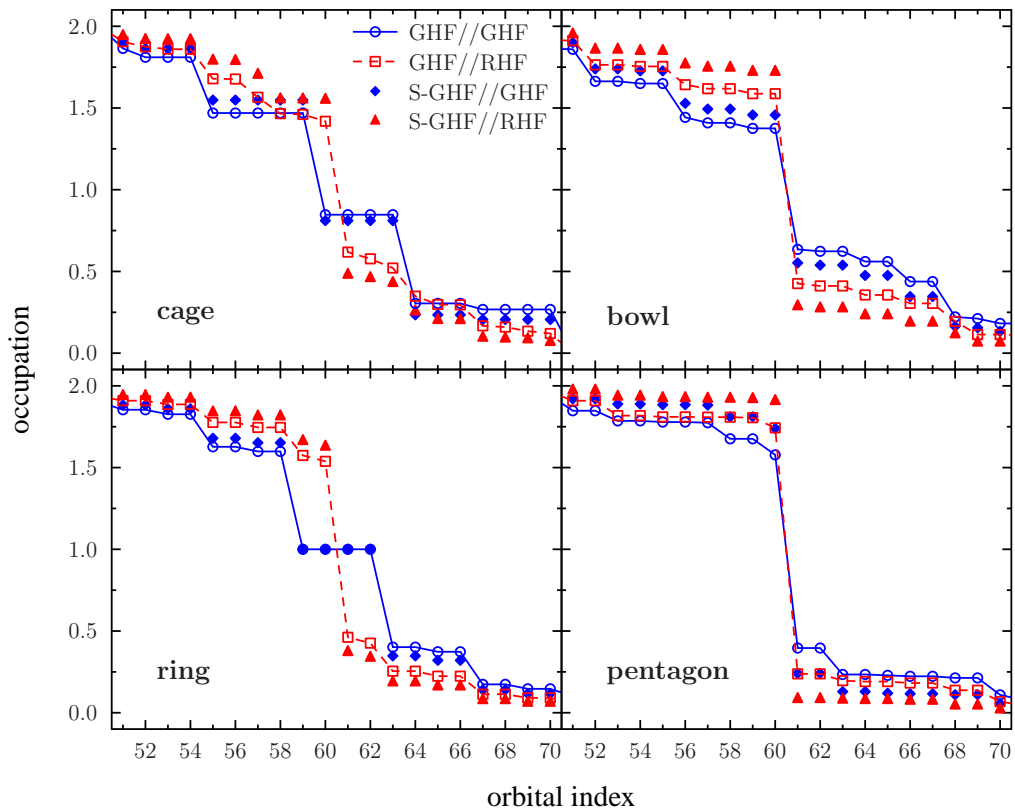}
  \caption{Natural occupations predicted by S-GHF ($s=0$) and charge
    natural occupations predicted by GHF for each of the four C$_{20}$
    isomers considered in this work, at the optimized RHF and GHF
    geometries. The 6-31G($d$) basis was used; only the occupations in
    the $\pi$-space are depicted. \label{fig:c20-occ}}
\end{figure}

We present in Fig. \ref{fig:c20-occ} the charge natural occupations
obtained by diagonalizing the charge density matrix $P_{\nu\mu} =
\sum_{s} \gamma^1_{\nu s, \mu s}$ associated with the GHF solution for
each of the four C$_{20}$ isomers under consideration. This is
equivalent to the well-known procedure to obtain the natural orbitals
and natural occupations from UHF wave functions
\cite{Pulay-1988}. Additionally, we present the natural occupations
(as true eigenvalues of the density matrix) obtained after carrying
out the (singlet) spin-projection from the GHF wave function.

From Fig. \ref{fig:c20-occ}, one immediately realizes that the GHF
charge natural occupations become closer to one at the GHF geometries
(as compared to the RHF ones). This is a consequence of the bond
length equalization observed in GHF optimized geometries, as opposed
to the alternating double- and single-bond character predicted by
RHF. It is also evident from the figure that S-GHF reduces the amount
of static correlation predicted by GHF, as occupation numbers become
closer to zero and two with the former, more accurate method. The
inclusion of dynamical correlation may contribute to this effect even
further. Aside from these generic trends, we see marked differences
between the four C$_{20}$ isomers. The C$_{20}$ pentagon displays few
signatures of static correlation, and may be accurately treated by
single-reference based methods. The bowl, cage, and ring isomers do
display features of static correlation, with several occupations
between $0.4$ and $1.6$. An accurate description of the electron
correlation in such systems would require either a multi-reference or
a symmetry-broken based treatment. In particular, the ring isomer
displays four occupations $\approx 1$ (they become exactly one for a
$D_{20h}$ geometry). These orbitals are thus singly occupied, though
entangled in an overall singlet wave function for S-GHF.

\begin{figure}
  \includegraphics[width=9.3cm]{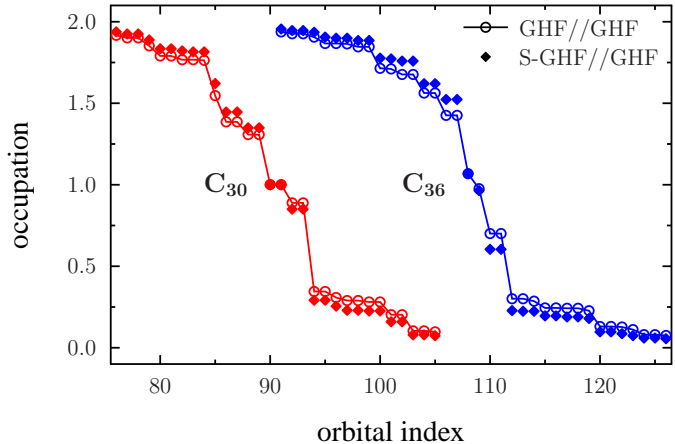}
  \caption{Same as Fig. \ref{fig:c20-occ}, but for the C$_{30}$ and
    C$_{36}$ fullerene cages. \label{fig:large-occ}}
\end{figure}

We show in Fig. \ref{fig:large-occ} the charge natural occupations
predicted by GHF and the natural occupations predicted by S-GHF
($s=0$) for C$_{30}$ and C$_{36}$, evaluated at the optimized GHF
geometries. As observed in the isomers of C$_{20}$, S-GHF reduces the
apparent amount of static correlation predicted by GHF, bringing the
occupations closer to 0 or 2 (albeit only slightly). Both C$_{30}$ and
C$_{36}$ can be immediately characterized as possessing large static
correlation, as already observed in the literature for C$_{36}$
\cite{Fowler-1999,Varganov-2002}. In particular, they both display a
few occupations near 1.

\begin{figure}
  \includegraphics[width=10.8cm]{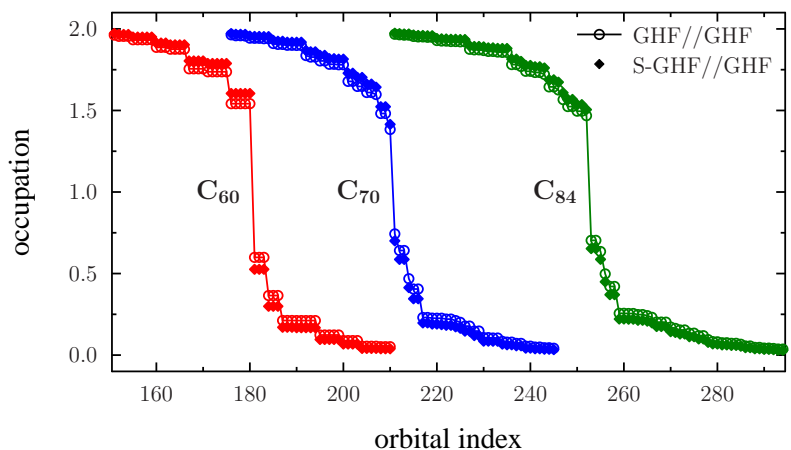}
  \caption{Same as Fig. \ref{fig:c20-occ}, but for the C$_{60}$,
    C$_{70}$, and C$_{84}$ fullerene cages. \label{fig:large-occ2}}
\end{figure}

Fig. \ref{fig:large-occ2} shows the natural occupations in C$_{60}$,
C$_{70}$, and C$_{84}$. In all of these cases we observe a significant
gap in the occupation numbers, although the gap is reduced for the two
larger cages. In C$_{60}$, for instance, S-GHF predicts no occupation
in the range $0.6<n<1.4$. Nevertheless, we would argue that the
profiles are still far from being single-determinantal in nature. We
note that the C$_{60}$ GHF occupation profile displays more static
correlation than that reported with UHF by St\"uck {\em et al.}
\cite{Stuck-2011} as there are more orbitals with occupation in the
range $0.4<n<1.6$.  We anticipate that single-reference based methods
(especially of the coupled-cluster type), can be significantly more
successful in C$_{60}$, C$_{70}$, or C$_{84}$ than in either C$_{30}$
or C$_{36}$.

Overall, the occupation profiles displayed for the fullerene isomers
are consistent with a polyradical character. The effective number of
unpaired electrons can be quantified using empirical formulas such as
that proposed by Head-Gordon \cite{HeadGordon-2003},
\begin{equation}
  N_{\textrm{unpaired}} = \sum_i \min(n_i,2-n_i).
\end{equation}
S-GHF/6-31G($d$) calculations, at the optimized GHF geometries,
predicts $10.2$, $8.8$, and $9.9$ unpaired electrons in C$_{30}$,
C$_{36}$, and C$_{60}$, respectively. This increases to $12.4$ and
$14.5$ for C$_{70}$ and C$_{84}$. The large polyradical character
displayed by the fullerenes rationalizes their increased reactivity as
compared to typical aromatic systems.

\subsection{Spin-spin correlations}
\label{sec:spin}

Having observed that GHF predicts magnetic moments in each carbon
center and that such structure is intrinsically connected with the
development of static correlation, the next thing is to establish the
nature of the interactions between the magnetic moments. In
particular, we study the inter-atomic spin-spin correlations using
Eq. \ref{eq:spin-spin}.

\begin{figure}
  \includegraphics[width=6cm]{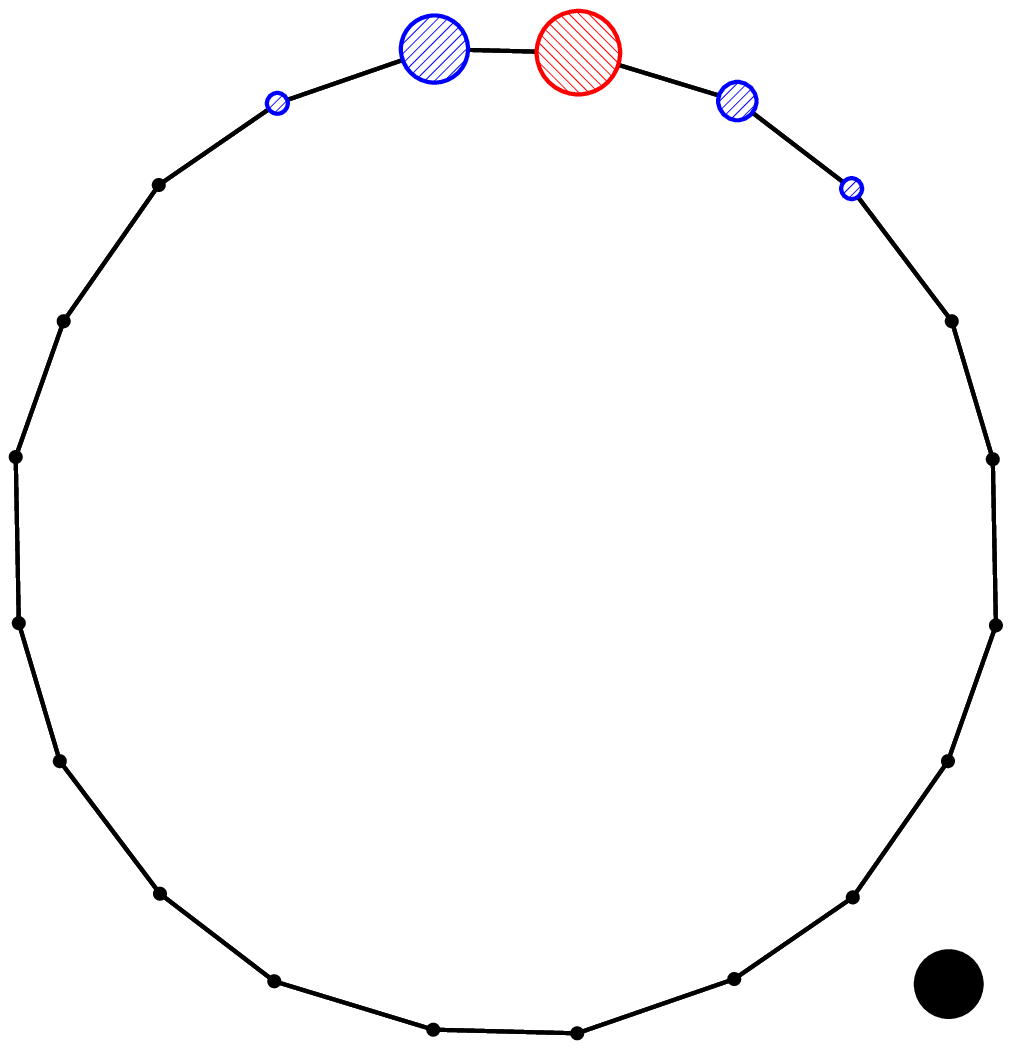}
  \hspace{0.5cm} \includegraphics[width=6cm]{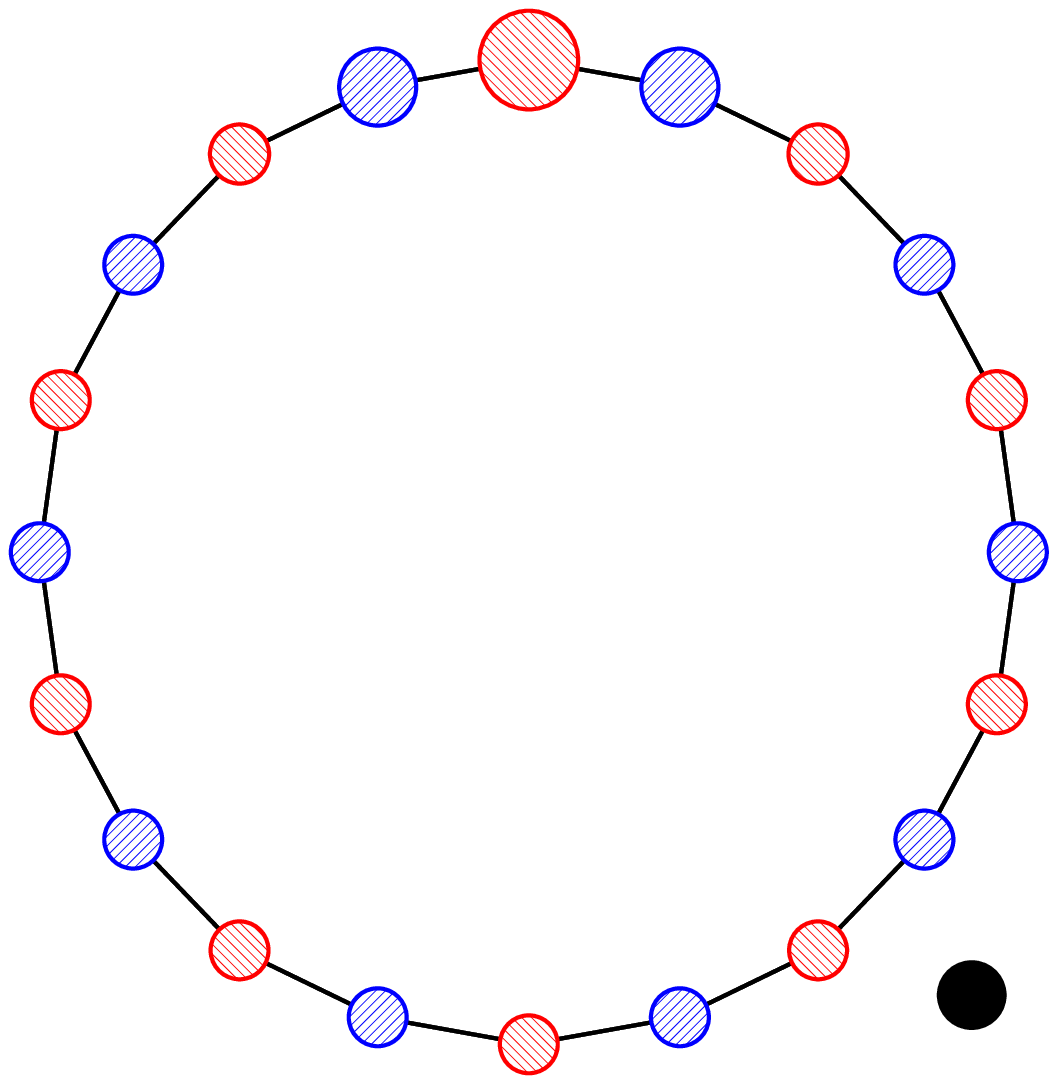}
  \caption{Atomic spin-spin correlations in the RHF/6-31G($d$) (left
    panel) and GHF/6-31G($d$) (right panel) solutions of the C$_{20}$
    ring, at the corresponding optimized geometries. A red (blue)
    circle indicates a positive (negative) correlation; the {\em area}
    of the circle is proportional to the magnitude of the
    correlation. The atom with the largest red circle (near the top of
    the figure) is used as probe. The black circle located at the
    bottom right of each figure is used to set the scale
    ($\mathbf{S}_A \cdot \mathbf{S}_B = 1$). \label{fig:c20r-spin}}
\end{figure}

We begin by considering the C$_{20}$ ring, for which GHF predicts an
axial spin-density wave character. We show in Fig. \ref{fig:c20r-spin}
the atomic spin-spin correlations predicted by RHF/6-31G($d$) and
GHF/6-31G($d$) at the corresponding optimized geometries. In RHF, the
magnetic moment (or spin) in each atom is strongly (negatively)
correlated with its doubly-bonded neighbor, and weakly with its
singly-bonded neighbor. Note that the RHF structure does not follow an
anti-ferromagnetic pattern. GHF, on the other hand, displays
long-range spin-spin correlations, as shown in the right panel. That
is, if the spin in the probe atom is up, one can know {\em a priori}
the spins in all other atoms, due to the anti-ferromagnetic character
of the spin density wave. This type of structure can be explained in
terms of strong nearest-neighbor-type spin-spin interactions between
the singly occupied $p$ orbital in each carbon atom, exactly those
considered in model Hamiltonian approaches.

We note that only the relative magnitudes of the atomic spin-spin
correlations should be given physical meaning. The fact that the
on-site spin-spin correlation is significantly larger than $3/4$ can
be attributed to contributions from the $\sigma$-bonded electrons.

\begin{figure}
  \includegraphics[width=6cm]{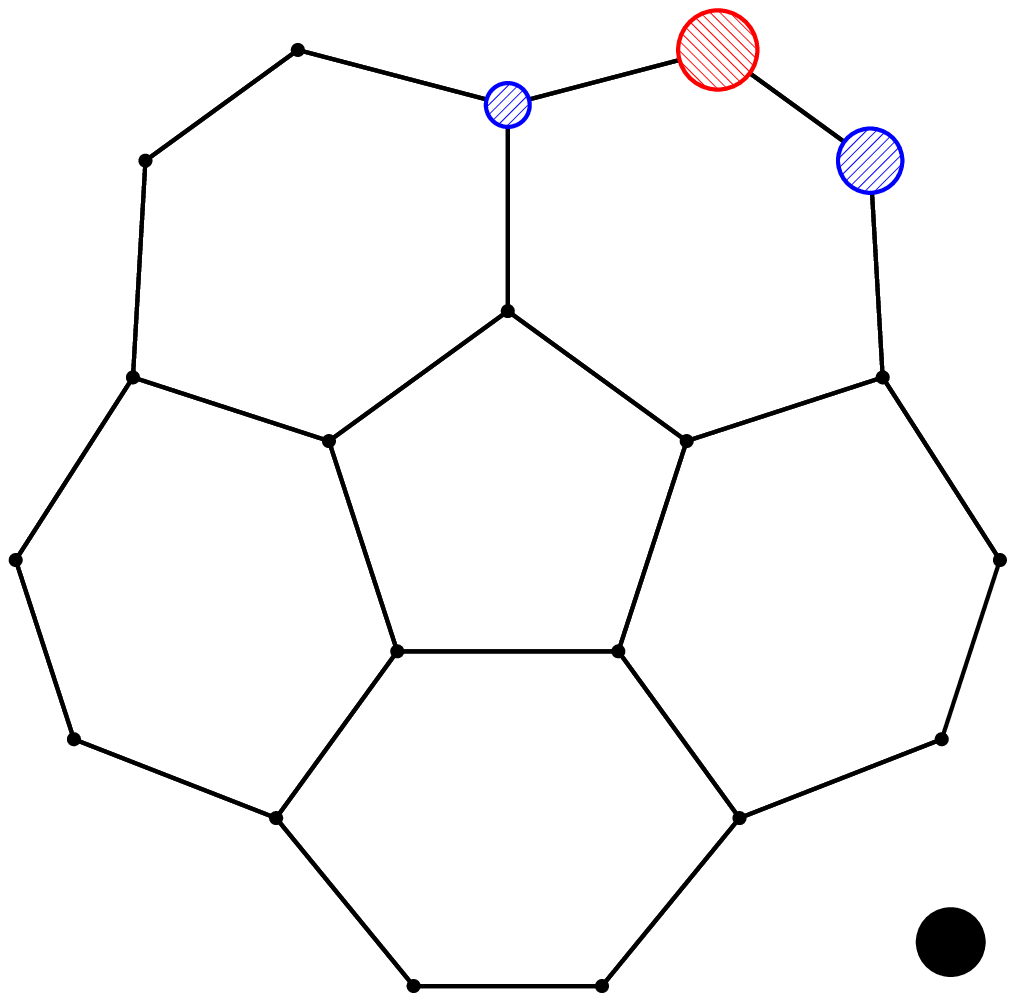}
  \hspace{0.5cm} \includegraphics[width=6cm]{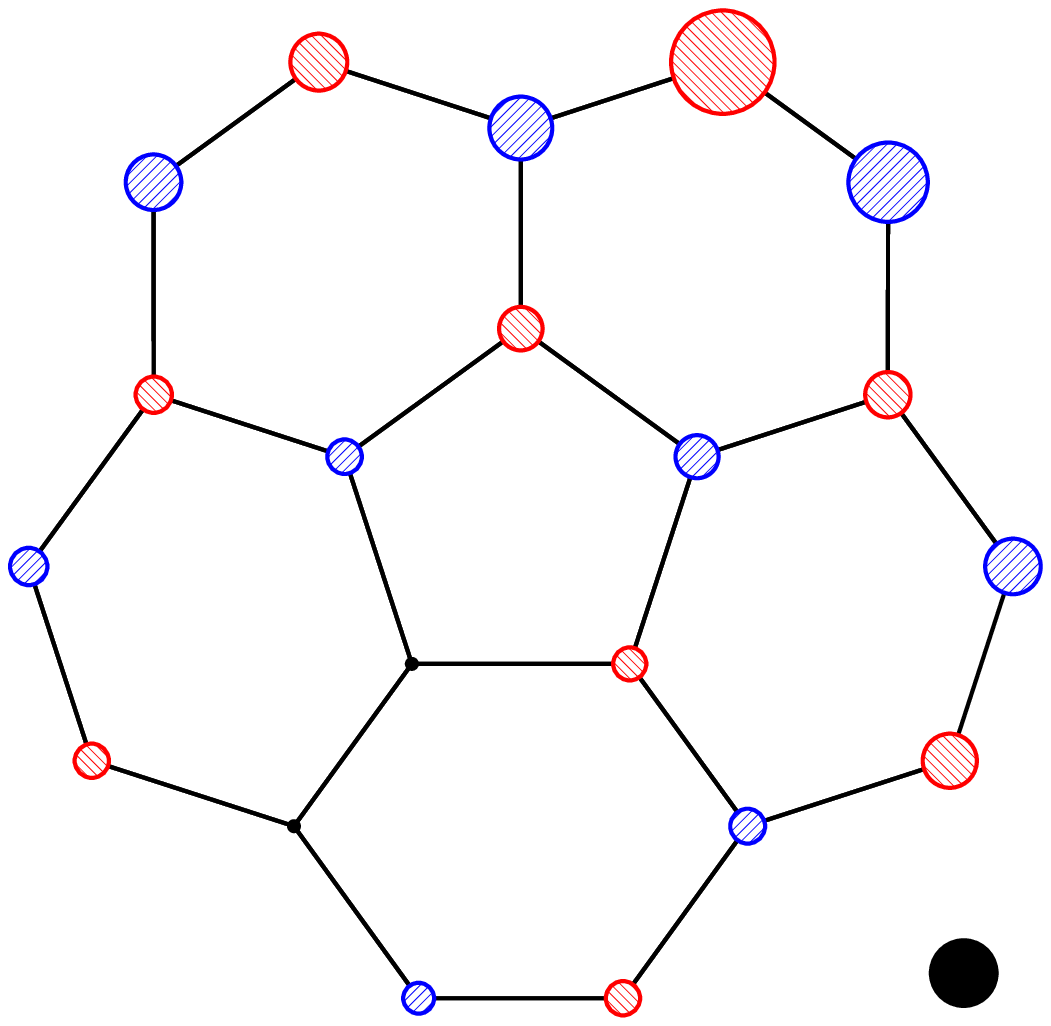}
  \caption{Same as Fig. \ref{fig:c20r-spin}, but for the bowl isomer
    of C$_{20}$; the probe atoms are located near the top-right corner
    of the figure. The left panel corresponds to the RHF solution,
    while the right displays the GHF solution. \label{fig:c20b-spin}}
\end{figure}

The same type of behavior is observed in other isomers. We show in
Fig. \ref{fig:c20b-spin} the atomic spin-spin correlations in the
C$_{20}$ bowl isomer, using an atom in the edge as a probe. Again, RHF
predicts strong correlations only with its triply-bonded neighbor. On
the other hand, GHF predicts strong correlations of long-range
character.

\begin{figure}
  \includegraphics[width=6cm]{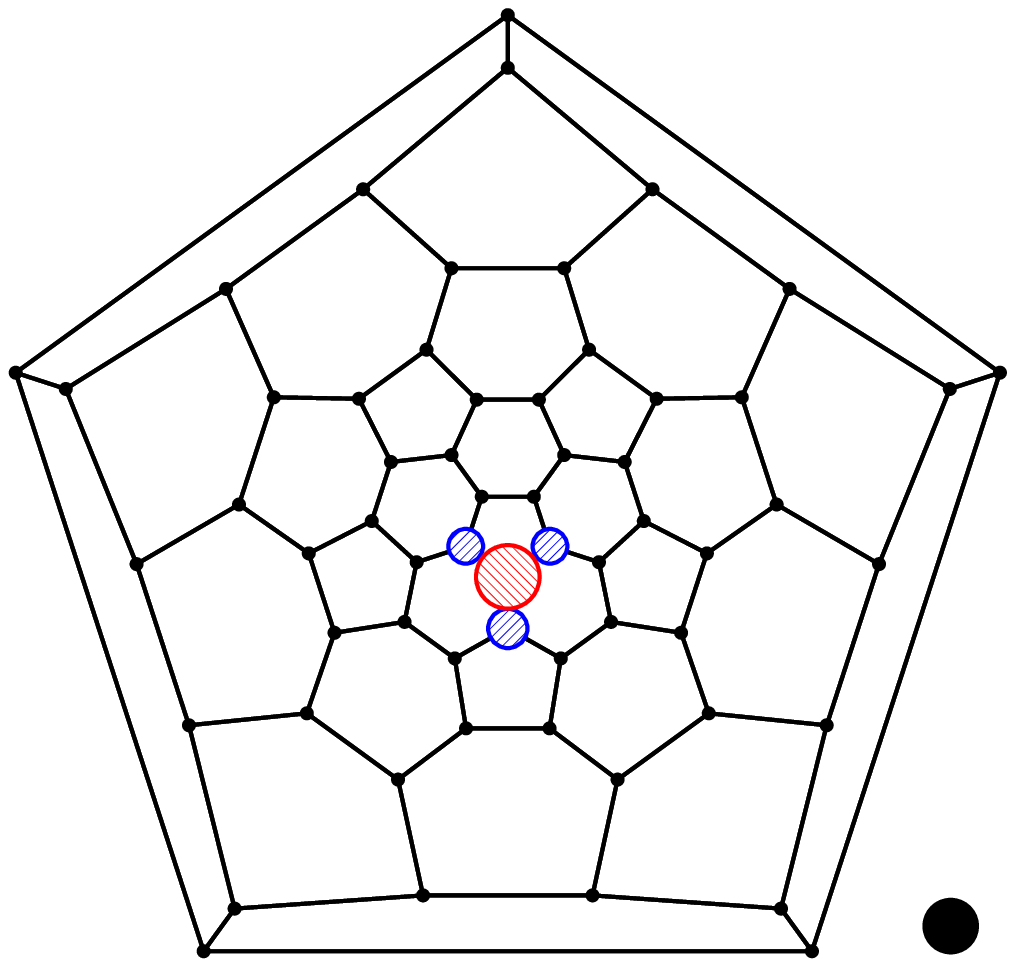}
  \hspace{0.5cm} \includegraphics[width=6cm]{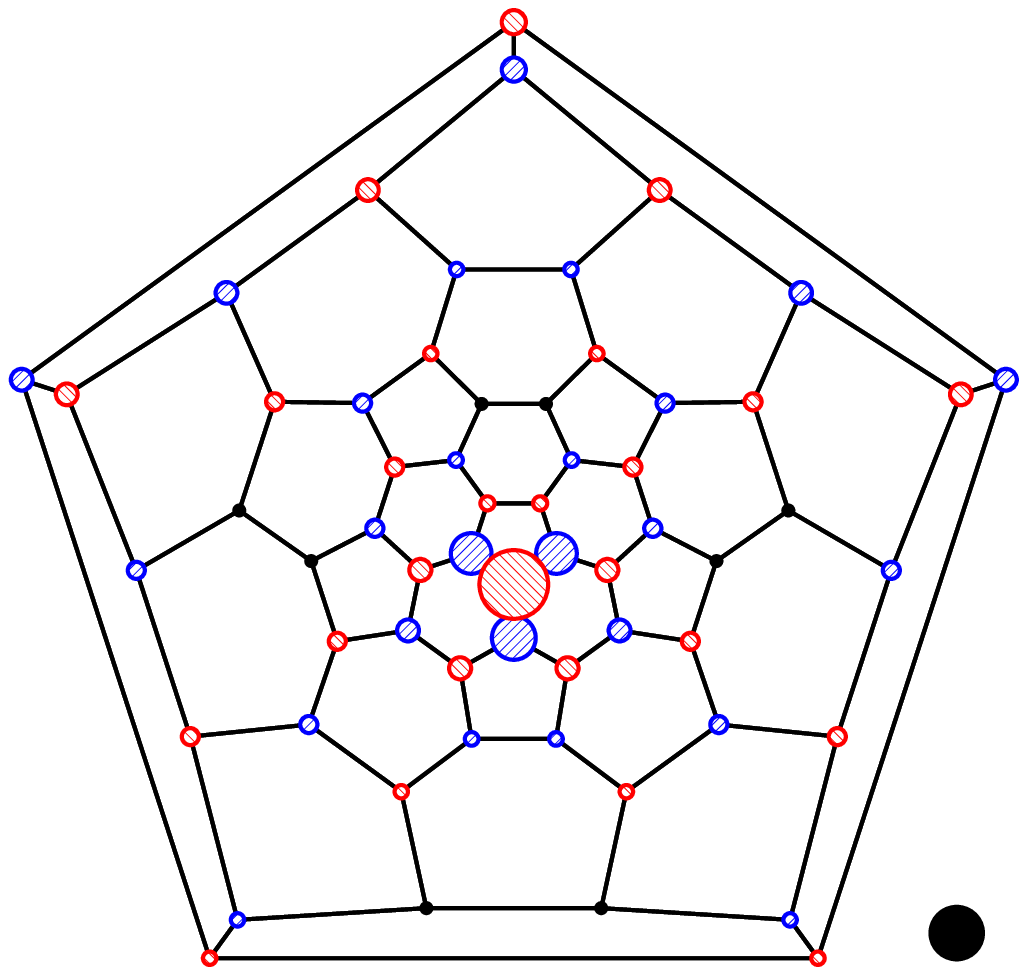}
  \caption{Same as Fig. \ref{fig:c20r-spin}, but for the C$_{60}$
    fullerene cage shown as a Schlegel diagram, with the probe atoms
    located near the center of the structure. The left panel
    corresponds to the RHF solution, while the right displays the GHF
    solution. \label{fig:c60-spin}}
\end{figure}

To show that GHF still predicts a strongly correlated nature between
localized spins in C$_{60}$, we show in Fig. \ref{fig:c60-spin} the
atomic spin-spin correlations in this molecule predicted by RHF and
GHF. The RHF solution predicts that there are significant spin-spin
correlations between the carbon center and its three nearest
neighbors. GHF yields the same type of strong correlations, but it
also displays non-vanishing spin-spin correlations with most other
atoms in the fullerene lattice. The nature of the GHF electronic wave
function is thus polyradical in character. We note that the ratio of
the magnitudes of the spin-spin correlations between the 6-5 and 6-6
bonds is $0.86$, significantly larger than the ratio obtained from
accurate solutions to the Heisenberg Hamiltonian \cite{Flocke-1998}. A
large part of this effect is, nevertheless, due to contributions from
$\sigma$-bonded electrons.

\begin{figure}
  \includegraphics[width=6cm]{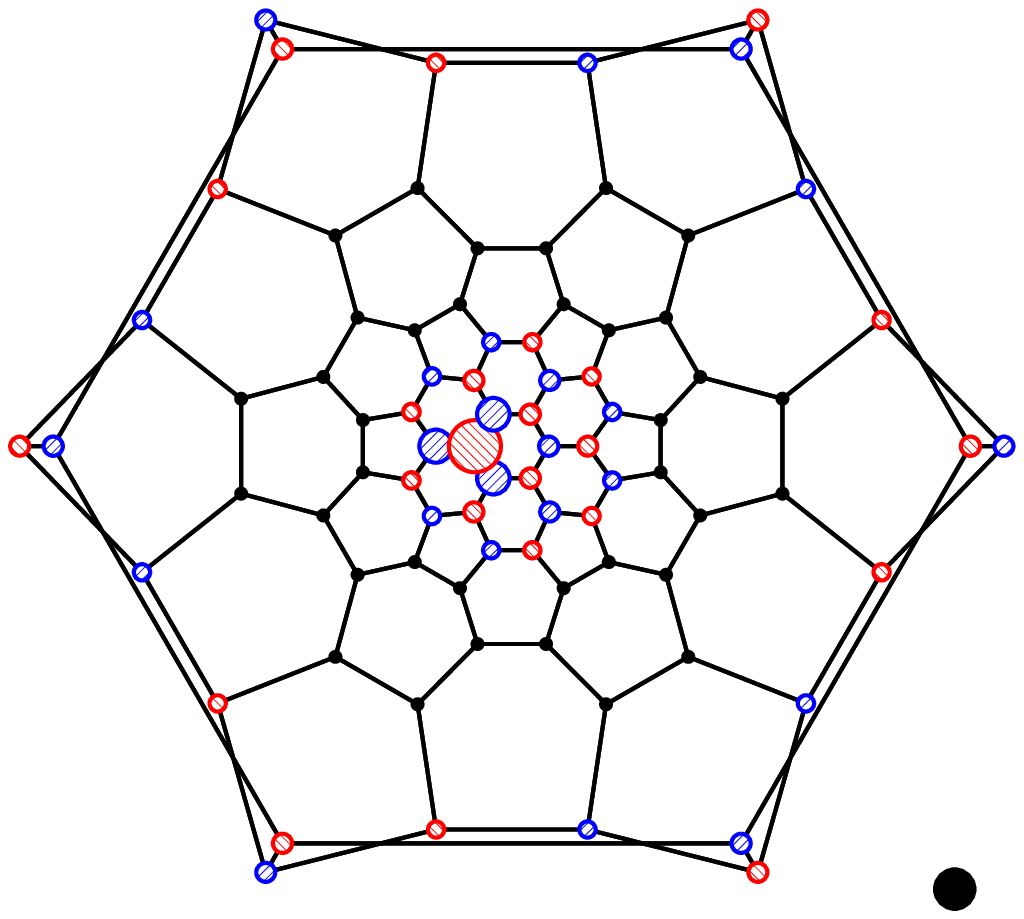}
  \hspace{0.5cm} \includegraphics[width=6cm]{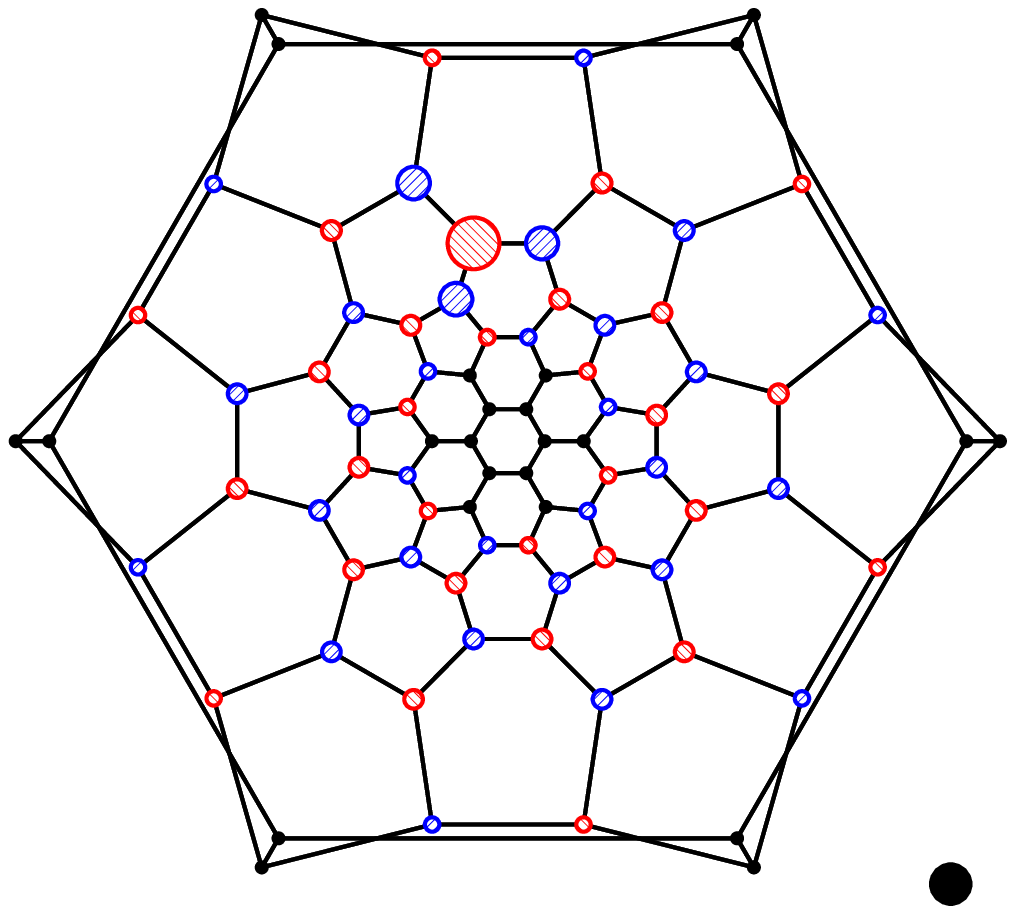}
  \caption{Same as Fig. \ref{fig:c20r-spin}, but for the C$_{84}$
    ($D_{6h}$) fullerene cage shown as a Schlegel diagram. Only the
    GHF solution is displayed, with the probe atom located in the pole
    (left panel) or in the equator (right panel) of the
    molecule. \label{fig:c84-spin}}
\end{figure}

For C$_{70}$ (not shown) and C$_{84}$ (see Fig. \ref{fig:c84-spin}),
we again observe the development of long-range anti-ferromagnetic
ordering. Interestingly, it appears that the atomic spin-spin
correlations are separately enhanced in the equator and the poles of
the molecules.

In Ref. \cite{Stuck-2011}, St\"uck {\em et al.} concluded that there
is no strong correlation present in C$_{60}$ according to the
occupation profile predicted by UHF and optimized scaled opposite-spin
MP2, and the fact that RHF-based MP2 predicted a more accurate
singlet-triplet gap than UHF-based calculations. They interpreted the
spin symmetry breaking occurring in C$_{60}$ as due to relatively
small but global electron correlations in the $\pi$-space, in spite of
the large spin density with anti-ferromagnetic character. Our results
support a different interpretation, one where the system is strongly
correlated in the sense that short-range spin-spin interactions drive
long-range magnetic ordering. This is more consistent with the nature
of valence bond solutions to the Heisenberg Hamiltonian
\cite{Schmalz-2002} and with accurate Monte-Carlo results to the
Hubbard Hamiltonian (see, {\em e.g.}, Ref. \cite{Scalettar-1993}). GHF
predicts the correct qualitative physical picture, though quantitative
agreement with the exact solution can only be obtained with correlated
approaches.

\subsection{Discussion}
\label{sec:disc}

Previous works have related the observed large spin contamination in
independent-particle model wave functions (HF and DFT) to the concept
of polyradicalism \cite{Bendikov-2004}. At this point, this relation
seems transparent in fullerene and related systems (such as
polyacenes): the large spin contamination in symmetry broken
approaches appears due to large atomic spin-spin correlations. The
latter are in turn a result of spins being partially localized in the
carbon centers so as to maximize anti-ferromagnetic interactions
(including minimizing the inherent frustration present in the system).

One must ask whether the physical picture portrayed by these solutions
is correct. We find three problems with GHF solutions that one must
keep in mind:
\begin{itemize}
  \item GHF predicts non-zero atomic magnetic moments on the atoms,
    with the spins aligned according to a pattern that minimizes
    frustration. This is an {\em unphysical} effect for a true singlet
    solution, where the average net spin density should identically
    vanish over all space. In other words, the correct description
    should have {\em fluctuating} spins, as opposed to the {\em
      permanent} spins in broken symmetry approaches, that yield a net
    zero magnetization at every point in space.

  \item The GHF solution displays a large spin contamination. In other
    words, the sum over the atomic spin-spin correlations results in a
    non-zero value for a singlet state.

  \item The broken-symmetry GHF solution predicts long-range ordering
    of the spins. The magnitude of the spin-spin correlation is
    roughly the same for, say, the third neighbor than for a far-away
    neighbor, as seen for the C$_{20}$ ring.
\end{itemize}
The first two problems can be easily fixed by spin-projection of the
optimized GHF wave function. S-GHF predicts a zero atomic magnetic
moment on each atom, and the sum of the inter-atomic spin-spin
correlations identically vanishes for singlet states.

\begin{figure}
  \includegraphics[width=13.8cm]{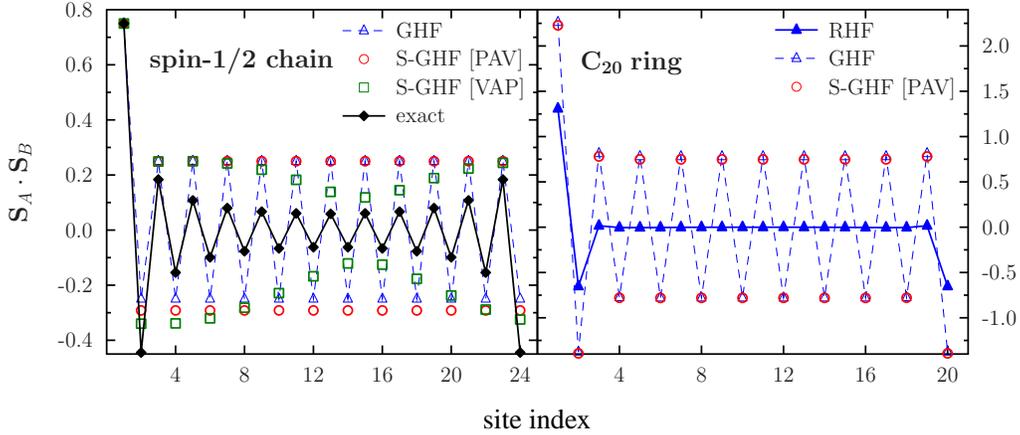}
  \caption{Spin-spin correlations predicted by HF and
    symmetry-projected HF in a 24-site spin-$1/2$ periodic Heisenberg
    chain (left) and in the C$_{20}$ ring (right). For the C$_{20}$
    ring, the calculations use the 6-31G($d$) basis set and were
    carried out at the GHF/6-31G($d$) optimized geometries. PAV and
    VAP denote projection-after-variation and
    variation-after-projection calculations. The latter implies a
    re-optimization of the orbitals in the presence of the projection
    operator. The exact results for the spin-$1/2$ periodic chain are
    from Ref. \cite{Lin-1991}. \label{fig:spin-spin}}
\end{figure}

The third effect is more subtle, although broken symmetry HF tends to
overemphasize the long-range correlations. We illustrate this by
comparing the spin-spin correlations in the C$_{20}$ ring with those
from a $24$-site periodic $s=1/2$ Heisenberg chain, as shown in
Fig. \ref{fig:spin-spin}. Whereas RHF fails to produce any type of
magnetic ordering, GHF overemphasizes the strength of the long-range
spin-spin correlations. The spin-spin correlations are known to decay
as $\sim \log^\sigma r/r$ in one-dimensional spin chains, with $\sigma
\approx 0.5$ \cite{Lin-1991,Mikeska-2004}. The spin-spin correlations
also display long-range ordering with a power law decay in
one-dimensional Hubbard chains (see, {\em e.g.},
Refs. \cite{Essler,RodriguezGuzman-2013}). It is clear from the figure
that a simple projection-after-variation spin projection does not
change the long-range character of the GHF correlations; it simply
reweighs some of the values so that the sum adds up to zero (for a
singlet state). We note that if a re-optimization of the orbitals is
allowed (in a variation-after-projection scheme), S-GHF may change the
structure of the spin-spin correlations by introducing defects into
the Slater determinant (see Fig. \ref{fig:spin-spin}).

For highly symmetric systems, one can gain a clearer insight to the
inter-atomic spin-spin correlations by decomposing them into the
different irreducible components. This is commonly done in periodic
systems and was also done by Srinivasan and co-workers
\cite{Srinivasan-1996} for C$_{60}$, which lead them to conclude that
the spin structure has large weights in the $T_{2g}$ and $G_u$
representations. We find it more convenient to introduce simple
measures of the strength of the overall spin-spin correlations and the
``long-range'' part of them. Namely, we introduce the following
measures:
\begin{align}
  \langle S^2 \rangle
    &= \, \sum_{AB} \mathbf{S}_A \cdot \mathbf{S}_B,
    \label{eq:s2} \\
  \langle S^2 \rangle_{\textrm{abs}}
    &= \, \frac{1}{N_A} \sum_{AB} |\mathbf{S}_A \cdot \mathbf{S}_B|,
    \label{eq:s2-abs} \\
  \langle S^2 \rangle_{\textrm{abs,lr}}
    &= \, \langle S^2 \rangle_{\textrm{abs}}
      - \frac{1}{N_A} \sum_{A} |\mathbf{S}_A \cdot \mathbf{S}_A|
      - \frac{2}{N_A} \sum_{<AB>} |\mathbf{S}_A \cdot \mathbf{S}_B|.
    \label{eq:s2-abs-lr}
\end{align}
Whereas $\langle S^2 \rangle$ should sum to zero for a singlet state,
$\langle S^2 \rangle_{\textrm{abs}}$ measures the overall strength of
the spin-spin correlations by summing the absolute values of the
inter-atomic interactions. The normalization factor $1/N_A$ is
introduced for convenience. If long-range order is present in a class
of systems, we expect $\langle S^2 \rangle_{\textrm{abs}}$ to scale
linearly with the size of them. $\langle S^2 \rangle_{\textrm{abs,lr}}$
measures the ``long-range'' part of the interactions by subtracting
the on-site and the nearest-neighbor correlations from $\langle S^2
\rangle_{\textrm{abs}}$.

We have computed these three quantities for the systems under
consideration; the results are presented in Table
\ref{tab:spin}. Neither RHF nor S-GHF present any spin
contamination. The RHF wave functions introduce almost no long-range
spin-spin correlations, as we had pointed out before in the figures,
and give almost constant values for $\langle S^2
\rangle_{\textrm{abs}}$. On the other hand, both GHF and its
spin-projected version introduce magnetic ordering in the systems
which becomes reflected in large $\langle S^2 \rangle_{\textrm{abs}}$
and $\langle S^2 \rangle_{\textrm{abs,lr}}$ values. The largest values
observed are for the C$_{20}$ ring, which is unsurprising given the
axial character of the spin density wave. For the fullerene cages,
$\langle S^2 \rangle_{\textrm{abs}}$ and its long-range counterpart
increase with the size of the system, but they do so in a sub-linear
way. This can be understood from the fact that the geometric
structures of the cages have different nature (in terms of adjacent
pentagons and hexagons) and the frustration included in the various
cages has different character.

\begin{table}
  \caption{Computed measures of the strength of spin-spin correlations
    as given by Eqs. \ref{eq:s2}, \ref{eq:s2-abs}, and
    \ref{eq:s2-abs-lr} for different systems. The 6-31G($d$) basis set
    was used. \label{tab:spin}}
    {\small
    \begin{tabular}{l r r r | r r r | r r r}
      \hline \hline
      & \multicolumn{3}{r}{RHF//RHF}
      & \multicolumn{3}{r}{GHF//GHF}
      & \multicolumn{3}{r}{S-GHF//GHF}
      \\[4pt] \cline{2-4} \cline{5-7} \cline{8-10}
      \\[-10pt]
      system
        & $\langle S^2 \rangle$
        & $\langle S^2 \rangle_{\textrm{abs}}$
        & $\langle S^2 \rangle_{\textrm{abs,lr}}$
        & $\langle S^2 \rangle$
        & $\langle S^2 \rangle_{\textrm{abs}}$
        & $\langle S^2 \rangle_{\textrm{abs,lr}}$
        & $\langle S^2 \rangle$
        & $\langle S^2 \rangle_{\textrm{abs}}$
        & $\langle S^2 \rangle_{\textrm{abs,lr}}$ \\[2pt]
      \hline
      C$_{20}$ bowl &
          0.00 &  2.97 &  0.05 &
          7.70 &  9.42 &  4.92 &
          0.00 &  9.40 &  4.85 \\[2pt]
      C$_{20}$ ring &
          0.00 &  3.02 &  0.16 &
          7.23 & 18.33 & 13.22 &
          0.00 & 18.05 & 12.94 \\[2pt]
      C$_{20}$ cage &
          0.00 &  2.91 &  0.09 &
          6.42 &  5.88 &  2.21 &
          0.00 &  5.91 &  2.18 \\[2pt]
      C$_{30}$ cage &
          0.00 &  2.91 &  0.08 &
          8.19 &  7.76 &  4.03 &
          0.00 &  7.71 &  3.96 \\[2pt]
      C$_{36}$ cage &
          0.00 &  2.96 &  0.10 &
          8.12 &  8.53 &  4.85 &
          0.00 &  8.35 &  4.66 \\[2pt]
      C$_{60}$ cage &
          0.00 &  2.98 &  0.08 &
         10.26 &  9.22 &  5.61 &
          0.00 &  8.98 &  5.38 \\[2pt]
      C$_{70}$ cage &
          0.00 &  2.98 &  0.08 &
         12.09 & 11.20 &  7.54 &
          0.00 & 10.94 &  7.29 \\[2pt]
      C$_{84}$ cage &
          0.00 &  2.99 &  0.08 &
         13.99 & 14.46 & 10.78 &
          0.00 & 14.12 & 10.45 \\[2pt]
      \hline \hline
    \end{tabular}}
\end{table}

At this point, we can state what one expects from highly accurate
correlated calculations. $\langle S^2 \rangle_{\textrm{abs}}$ should
be much larger than in RHF but is likely going to be significantly
smaller than in GHF. The near zero value predicted by RHF for $\langle
S^2 \rangle_{\textrm{abs,lr}}$ is most certainly qualitatively
incorrect. Indeed, highly accurate quantum Monte Carlo calculations
\cite{Scalettar-1993} in the C$_{60}$ Hubbard lattice have shown that
spin correlations vanish at {\em large} distances, with a correlation
length of 3-4 bonds. Density-matrix renormalization group (DMRG)
calculations of Hachmann and coworkers in polyacene systems
\cite{Hachmann-2007} have also displayed magnetic ordering but with
decaying spin-spin correlations.

The nature of the chemical bond in large conjugated molecules becomes
dominated by short-range spin-spin type interactions. This is in
contrast to the delocalized, non-interacting electron-gas picture
offered by H\"uckel theory. We agree with Hachmann {\em et al.}
\cite{Hachmann-2007} and Schmalz \cite{Schmalz-2002} on the fact that
the bonding is most simply expressed in terms of a valence bond
framework. Ionic contributions (zero and double occupations) in the
$\pi$-orbitals are diminished in favor of a resonating valence bond
structure between the localized electrons in each carbon center. The
system becomes a Mott insulator \cite{Mott-1949}, rather than
displaying metallic-like behavior. This is likely true regardless of
the shape of the molecule (fullerene, graphene-like ribbon, carbon
nanotube, etc.). A broken symmetry HF approach is consistent with this
framework, but predicts too much magnetic ordering.  Ionic
contributions are however, still significant, as can be seen from the
occupation profiles displayed above. Although the occupations are far
from zero or two, they are clearly also far from being all identical
to one. We thus expect that a Heisenberg-type Hamiltonian can only
offer a qualitative picture of the nature of the spin-spin
interactions. The Hubbard and the Pariser-Parr-Pople model
Hamiltonians are more appropriate to understand the electron
correlation in the $\pi$-space of conjugated systems such as the
fullerenes.

\begin{figure}
  \includegraphics[width=13.8cm]{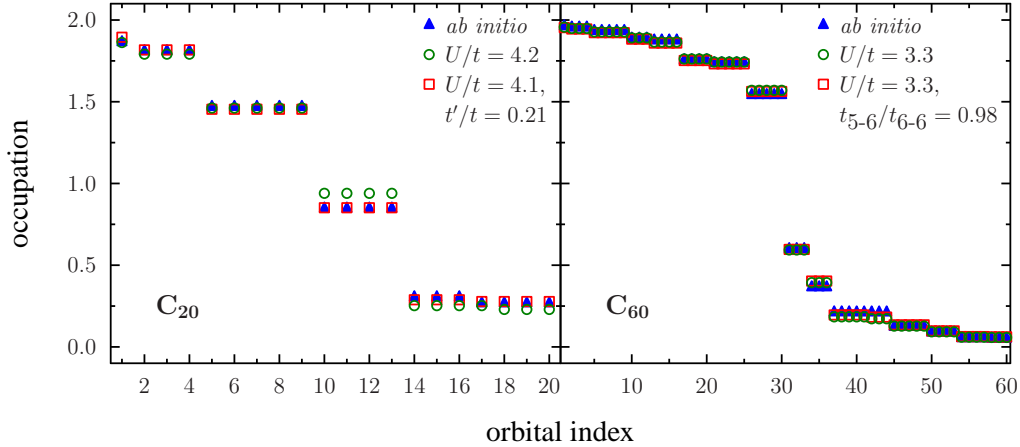}
  \caption{Comparison of GHF charge natural occupations predicted from
    the {\em ab initio} results and Hubbard Hamiltonian
    parametrizations for the C$_{20}$ (left) and C$_{60}$ (right)
    fullerene cages. More details are provided in the main
    text. \label{fig:hub-occ}}
\end{figure}

We note that the charge natural occupation profiles obtained in GHF
calculations are possibly useful in parametrizing Hubbard-type
Hamiltonians for fullerenes. Here, the Hubbard Hamiltonian is written
in the form
\begin{equation}
  \hat{H} = -\sum_{ij} t_{ij} \sum_{s=\uparrow,\downarrow} \left(
  c^\dagger_{i,s} \, c_{j,s} + c^\dagger_{j,s} \, c_{i,s} \right) + U
  \sum_i n_{i,\uparrow} \, n_{i,\downarrow},
\end{equation}
where $i,j$ label lattice sites in the fullerene, and $t$ and $U$
correspond to hopping (kinetic energy) and on-site (Coulomb) repulsion
amplitudes, respectively.  In particular, the charge natural
occupations predicted by GHF approximations in both the Hubbard
Hamiltonian and the {\em ab initio} system can be used in a
least-squares fit sense (see Fig. \ref{fig:hub-occ}). The fact that
the Hubbard Hamiltonian parameters are adjusted so that an approximate
method yields similar results is appealing.

The use of this strategy for the C$_{20}$ fullerene cage results in
$U/t \sim 4.2$ when only nearest-neighbor hopping is used. This places
the fullerene in the intermediate-to-strong correlated regime
(considering that the single-particle bandwidth is $\approx 6\,t$). As
shown in Fig. \ref{fig:hub-occ}, the GHF charge natural occupation
profile obtained from the Hubbard-type Hamiltonian resembles well the
{\em ab initio} one, save for the lowest-occupied states. A better fit
can be obtained using an additional next-nearest-neighbor hopping
amplitude $t'/t \sim 0.21$ and $U/t \sim 4.1$.

For the C$_{60}$ fullerene cage, assuming homogeneous hopping
parameters and only nearest-neighbor hopping allowed, one obtains $U/t
\sim 3.3$. As we have previously discussed, the electron correlation
in C$_{60}$ is weaker than in C$_{20}$, so this is unsurprising. The
charge natural occupation profile from the Hubbard Hamiltonian is in
fairly good agreement with the {\em ab initio} one. We obtain a
slightly better fit by allowing different hoppings in the 5-6 bonds
than in the 6-6 bonds: $t_{\textrm{5-6}}/t_{\textrm{6-6}} \sim 0.98$,
with $U/t \sim 3.3$.

We note that our $U/t$ value for C$_{60}$ is well within the range
$2\leq U/t \leq 5$ considered by other works
\cite{Bergomi-1993,Scalettar-1993,Joyes-1993,Sheng-1994,Krivnov-1994,Srinivasan-1996,Ojeda-1999,Lin-2007b}. Our
parametrization for $U/t$ for C$_{20}$ is significantly smaller than
that reported by Lin and S\o{}rensen \cite{Lin-2008}. The need for
next-nearest neighbor hopping in the latter lattice might be
understood in terms of the larger $s$ character of the hybridized
orbitals defining the $\pi$-space in C$_{20}$ than in C$_{60}$, due to
the higher curvature of the structure. We plan to study in the near
future the extent to which simple Hubbard-type parametrizations
reproduce the low-energy physics of selected fullerene cages.

\section{Conclusions}

In this work, we have shown a few examples of fullerene systems where
HF solutions of the generalized type (with non-collinear spin
arrangements) are lower in energy than their respective restricted
counterparts. In these solutions, effective localized magnetic moments
develop in the $p$ orbitals of each carbon center; the spins become
aligned in such a way so as to maximize anti-ferromagnetic
interactions and minimize frustration. The large spin contamination
observed in these solutions is indicative of strong inter-atomic
spin-spin correlations that should determine the low-energy physical
and chemical properties of fullerenes. The GHF-based optimized
geometries avoid Jahn-Teller type distortions in fullerene cages such
as C$_{20}$, C$_{30}$ or C$_{36}$. Fullerene molecules are predicted
to be polyradical in nature and thus either a multi-reference or a
broken symmetry treatment is necessary to get a correct qualitative
physical picture.

It would be interesting to tackle some of the examples here considered
with more accurate multi-reference based methods in order to obtain a
complete picture of the nature of the electronic structure in
fullerenes and of the importance of spin-spin correlations in
particular. Unfortunately, large active spaces seem inevitable (due to
the size of the systems and the fact that $\sigma$-orbitals cannot be
completely ignored).  This would also permit to assess the extent to
which model Hamiltonians describe the properties of fullerene systems
(see, {\em e.g.}, Refs. \cite{Pastor-1996,Schmalz-2002}) or,
equivalently, the extent to which the $\pi$-space is fully responsible
for low-energy electronic properties. We note that the radical
scavenger character of C$_{60}$ \cite{Taylor} suggests that spin-spin
correlations are indeed significant. An interesting alternative avenue
is to explore non-collinear density functional-based approaches for
the fullerenes (see, {\em e.g.}, Ref. \cite{Scalmani-2012}).

We think it would also be interesting to re-investigate some of the
chemical properties of fullerenes with GHF-based solutions as opposed
to the traditional restricted framework adopted either in HF or
density functional approaches. In particular, we are interested in the
extent to which spin-spin correlations account for the observed
addition patterns in fullerene reactions. Further investigations can
also explore how the spin structure is affected by incarcerated
species or by defects in the fullerene structure such as the
introduction of edges. The study of fullerene analogues such as
boron-nitride, silicon, or germanium-based structures is also
appealing.

Aside from that, we believe that fullerenes may become model systems
in which, on the one hand, both model Hamiltonian and full {\em ab
  initio} calculations can be carried out, and, on the other, actual
experiments can be performed. Given that doping is believed to be a
precursor for superconductivity in Mott-insulating environments, it
would be interesting to study the electronic structure of doped
systems, both by ionization and by chemical means. (We note that
chemists have been successful in preparing heterofullerenes
\cite{Taylor} where, {\em e.g.}, a carbon atom is replaced by an
electron-doped nitrogen atom or a hole-doped boron center.)

More generally, we believe that broken symmetry HF approaches can help
to understand the nature of the chemical bonding in other carbon-based
structures such as carbon nanotubes or graphene-like structures. The
recent work by Sheka and Chernozatonskii following this idea should be
highlighted \cite{Sheka-2007b,Sheka-2010,Sheka-2010b}. The spin
symmetry-projection strategy is important in order to eliminate the
unphysical effects in these solutions.

\begin{acknowledgement}

This work was supported by the National Science Foundation
(CHE-1110884). G.E.S. is a Welch Foundation Chair (C-0036). Some of
the computational resources used were provided in part by NIH award
NCRR S10RR02950 and an IBM Shared University Research (SUR) award in
partnership with CISCO, Qlogic and Adaptive Computing.

\end{acknowledgement}

\bibliography{fullerene}


\end{document}